\def\sun{$_\odot$}  
\documentstyle[12pt]{article}     

\def\alwaysmath#1{{\ifmmode{#1}\else{$#1$}\fi}}
\def\he#1{\hbox{\alwaysmath{{}^{#1}}{\rm He}}}

\def\li#1{\hbox{\alwaysmath{{}^{#1}}{\rm Li}}}

\def\etal{{\it et al.}~}
\def\beginapjbib{\begingroup \section*{\large \bf References}
   \parskip=.5ex plus 1.0pt
   \def\bibitem{\par \noindent \hangindent\parindent
      \hangafter=1}}
\def\endapjbib{\par \endgroup}
\def\endapjbib{\par \endgroup}
\def\beginfig{\begingroup \section*{\large \bf Figure Captions}
         \parskip=.5ex plus 1.0pt
         \def\figitim{\par \noindent \hangindent\parindent
                \hangafter=1}}
\def\endfig{\par \endgroup}
\begin{document}
\begin{titlepage}
\pagestyle{empty}
\baselineskip=21pt
\rightline{UMN-TH-1502/96}
\rightline{astro-ph/9607106}
\rightline{July 1996}
\vskip .2in
\begin{center}
{\large{\bf The Effects of an Early Galactic Wind  on  \\
the Evolution of D, \he3 and Z }}
\end{center}
\vskip .1in
\begin{center}

Sean Scully$^1$, Michel Cass\'{e} $^2$,
Keith A. Olive$^1$, 
and Elisabeth Vangioni-Flam$^3$

$^1${\it School of Physics and Astronomy, University of Minnesota}
{\it Minneapolis, MN 55455, USA}

$^2${\it Service d'Astrophysique, DSM, DAPNIA, CEA,  France}

$^3${\it Institut d'Astrophysique de Paris, 98bis
Boulevard Arago, 75014 Paris, France}

\vskip .1in

\end{center}
\vskip .5in
\centerline{ {\bf Abstract} }
\baselineskip=18pt

The predictions of the abundances of  D and \he3 from Big
Bang Nucleosynthesis (BBN) and recent observations of these two
isotopes suggest the need to develop new chemical evolution models. In 
particular, we examine the role of an early episode of massive star formation 
that would induce a strong destruction of D and a galactic wind. 
We discuss the ability of these models to match the observed local
properties of the solar neighborhood such as the
 gas mass fraction, oxygen abundance, the age-metallicity
relation, and the present-day mass
 function (PDMF).
We also examine in detail the ability of the chemical evolution models
 discussed  to reproduce the apparent lack of low mass, 
low metallicity stars in the solar neighborhood, namely the G-dwarf
 distribution. Indeed, we find models which satisfy the above constraints
while at the same time allowing  for a large primordial D/H ratio 
as is reportedly measured
in some quasar absorption systems at high $z$, without the overproduction
of heavy elements.  The latter constraint is achieved by employing a simple
dynamical model for a galactic wind.

\baselineskip=18pt

\noindent
\end{titlepage}
%\newpage
\baselineskip=18pt                   
\def\la{~\mbox{\raisebox{-.7ex}{$\stackrel{<}{\sim}$}}~}
\def\ga{~\mbox{\raisebox{-.7ex}{$\stackrel{>}{\sim}$}}~}
\def\beq{\begin{equation}}
\def\eeq{\end{equation}}

\section{Introduction}                  

In addition to the cosmic microwave background radiation and the Hubble
expansion, another testable prediction of the standard hot big bang model
is the synthesis of the light elements D, \he3, \he4, and,
\li7 (Walker \etal 1991). This prediction, tested against observations, 
is not always simple.
Most of these elements undergo significant galactic processing which has
changed their abundances over time. In the cases of \he4 and \li7,
primordial values can be reasonably well determined directly from observations.
\he4 may be inferred from low metallicity HII regions 
(see e.g. Pagel \etal 1992; Olive \& Steigman 1995; Olive \& Scully 1996).  
Observations of a uniform abundance of 
\li7 in halo dwarfs are interpreted to be the primordial value for
\li7/H (Spite \&
Spite 1982, Thorburn 1994 and Boesgaard 1996). Indeed, 
it has been argued that
 on the basis of these two isotopes.
one can confidently constrain the single parameter
  (the baryon-to-photon ratio, $\eta$) of  big bang nucleosynthesis
(Fields and Olive 1996, Fields \etal 1996).
On the other hand, in the cases of D and \he3, one observes
the present day abundances and solar abundances of these
elements. Therefore, an
understanding of the processes which alter the abundances of these elements
over time is necessary if one is to use the observational data 
to constrain their
primordial values as a test of BBN. The
evolution of D and \he3 will be examined here in terms of galactic
chemical evolution models.

In modeling the evolution of D/H, it is assumed that
all of the observed deuterium is primordial, as stars completely destroy
D in their pre-main-sequence phase.  As
a result, the present interstellar medium (ISM) abundances
 of D  are depleted from their primordial values and represent firm lower
bounds to the primordial abundance.  
The solar value of D/H can only be indirectly determined from the difference
 between the \he3/H ratio in the solar wind (which contains \he3 
from processed D) and in meteorites (Geiss 1993). The solar value of D/H
may also be determined from abundance measurement on the surface of 
Jupiter (Niemann \etal 1996, Ben Jaffel \etal 1996).
In addition, classical stellar models  predict 
(cf. Iben \& Truran 1978) that low mass stars 
are net producers of \he3, but
more massive stars destroy some fraction of it (Dearborn, Schramm \& Steigman,
1986).
Because of the pre-main-sequence destruction of D and its conversion to \he3, 
the evolution of these two elements is closely tied together. 
 The amount of \he3 ejected
from a star depends greatly on the amount of D present at its birth. 
Uncertainties, therefore, not only exist in the chemical evolution
but above all in the stellar production of \he3, as shown clearly by Schatzman
(1987).
 
Vangioni-Flam, Olive \& Prantzos (1994) explored the evolution of D
  and \he3
choosing primordial values of these elements corresponding to  baryon
to a photon ratio $\eta = 3 \times 10^{-10}$, a value consistent with the
primordial values inferred from observations of \he4 and \li7 and the canonical 
treatment of the evolution of D and \he3 (Walker \etal 1991).
Using closed box models of galactic chemical evolution, they found that
\he3 is overproduced compared with the observed solar and present-day
values of \he3 unless it is assumed that \he3 is destroyed
significantly  in low-mass stars
(ie, at levels comparable to the destruction in more 
massive stars).  What is problematic however, is that 
there is strong evidence that \he3 is produced in low mass stars since
abundances as high as $^3$He/H $\sim 10^{-3}$ have been observed 
in planetary nebulae (Rood, Bania
\& Wilson, 1992; Rood \etal 1995). Including the production of 
\he3 as calculated by Iben \& Truran (1978), however leads to a
substantial overproduction of \he3 at the solar epoch (Olive \etal 1995, 
Galli \etal 1995, Dearborn \etal 1996).
This problem may be remedied if
\he3 is only produced in a narrow mass range
(0.9 to 1.0 M$_\odot$), in agreement with the scarcity of 
\he3 rich planetary nebulae
(Rood \etal 1995) and destroyed in higher mass stars as we show below. 
For the most part, however, we will concentrate on the evolution of D/H and 
the metallicity as traced by O/H and Fe/H. As shown in Scully \etal (1996), even
in chemical evolution models with rather extreme assumptions regarding star 
formation rates and stellar and galactic winds, the problem
of \he3 can not be fully resolved.  Thus it seems evident a proper understanding
of \he3 evolution requires
some major modification in the stellar evolution of \he3.

To make matters more difficult (from the point of view of chemical evolution), 
some recent observations of D/H in quasar absorption systems indicate a high
primordial value around $2 \times 10^{-4}$ (Carswell \etal 1994; Songaila \etal
1994), requiring destruction factors  in excess of
$\sim$10. There remains some reason to be cautious of the original
observations because of  reports of D/H an order of magnitude lower in
other quasar absorption systems (Tytler, Fan, \& Burles 1996,
Burles \& Tytler 1996).  A recent
re-observation of the high D/H absorption system has been resolved in to 
two components, both yielding high values with an average value of D/H = $1.9
\pm 0.4 \times 10^{-4}$ (Rugers \& Hogan 1995a) as well as 
additional systems with a similar high value (Rugers \& Hogan 1996b,c). Other 
high D/H ratios were reported by Carswell \etal (1996) and Wampler \etal
(1996).

 The
cosmological consequences of a high primordial abundance of D/H
 was discussed by Vangioni-Flam and Cass\'{e} (1995) and by
Dar (1995). Interestingly enough,
 in a  recent
analysis of the predictions of BBN neglecting the observational data on solar
and ISM D and \he3 (justified by the current uncertainties in both the stellar
 and chemical evolution of these isotopes), Fields
\& Olive (1996) and Fields \etal (1996) found that the 
reasonably well determined abundances of \he4 and
\li7 lead to the prediction of a low value for $\eta = 1.8^{+1.0}_{-0.2} \times
10^{-10}$ which corresponds of a value of D/H = $1.8^{+0.5}_{-0.9} \times
10^{-4}$ in excellent agreement with the quasar absorption system measurements.
In contrast, the low D/H measurements, if they truly represent the primordial
value are not consistent with \he4 or \li7 unless it can be argued that
the observed \he4 abundances are seriously underestimated and that \li7
 has been significantly depleted.

Large deuterium destruction factors (of order 5 -- 10 from primordial values to
present) in models of galactic chemical evolution were first shown to be viable
by Vangioni-Flam \& Audouze (1988). These models typically assumed
an enhanced rate of massive star
formation in the early stages of galactic evolution. However, these
types of 
models have tended to
be criticized: Reeves (1991) argued that on the basis of the nuclear
chronometers, no more than a factor of 2 -- 3 destruction was possible, though
it was later shown in Scully \& Olive (1995) that the chronometer constraints
could in fact be  satisfied in a large class of models which provide for
significant D destruction. It is also often claimed (see eg. Edmunds 1995) that
models with large D destruction factors suffer from a G-dwarf problem. We will
clarify the situation with regard to the G-dwarf distribution below.
 For deuterium destruction factors above 10, 
other solutions have been proposed
 which invoke a galactic wind in the primitive galaxy
 (Vangioni Flam and Cass\'{e} 1995). 

A large deuterium destruction factor requires a high degree of gas
processing. This large
astration could, in turn, induce an excessive metallicity in the ISM. One
possibility to avoid the over-production of metals is to invoke 
an early galactic wind driven by supernovae (SN). Such a wind
 could keep the abundance of metals down to a reasonable level 
in the galactic disk and at the same
time remove gas from this reservoir. Furthermore, the D free ejecta of low mass,
long lived stars, diluted in relatively small amounts of the ISM would induce a
large decrease of the D/H ratio at late times.
Thus, the evolution of both D and Z can be strongly influenced by the presence
 of galactic winds or outflow (Vangioni-Flam \& Cass\'{e}, 1995).

 The possibility of a galactic wind which is driven directly by supernovae
  explosions is in fact well 
documented (Larson 1974, Vader 1986, Charlton \& 
 Salpeter 1989, David \etal
  1990, Wang \& Silk 1993).
In the following, we will couple a simple wind model inspired by Larson (1974)
 and David \etal (1990) to a chemical evolutionary code. Our
  criterion is the following: an early galactic wind can develop only if
  the supernova remnants overlap before radiatively dissipating all of
their energy.
 This ensures an efficient conversion of the explosive supernova energy into
  bulk kinetic energy of the surrounding medium. The maximum mass removed per
 supernova (neglecting radiative losses) is that accelerated above the escape
 velocity, about 500 km/s in the solar vicinity, and in the most
 favorable conditions the galactic mass loss rate is 
 $\sim$ 40 $\dot N_{SN}$ M$_\odot$, where $\dot N_{SN}$ 
is the rate of core collapse SN.   

To produce the outflow and the desired destruction of D, 
 it is necessary  to choose an IMF which is
skewed toward producing more massive stars, especially at early times.
These models have the advantage of quickly
 ejecting large amounts of D and \he3 free material and are able to
 generate winds in favorable conditions. However,  producing more
 massive stars will lead to generous heavy element production. 
 In particular, since these stars are the primary producers of
$^{16}$O and  produce a large amount of $^{56}$Fe as well, including more of
these stars will lead to an overproduction of these elements. Galactic winds of
the type discussed above is a way to reduce or remove 
the overproduction of metals in these models.  

Changing the IMF in favor of more massive stars may lead to a
discrepancy between the predicted and observed PDMF
(see e.g. Scalo 1986).
As a result, we consider the possibility that more massive stars
were formed early on in galactic history while star formation today
matches a more normal IMF.
 A class of models have been proposed in which the star formation is
bimodal (see e.g. Larson, 1986; Wyse and Silk, 1987; Vangioni-Flam and
Audouze, 1988; Fran\c{c}ois et al. 1990).  
In these models, the formation of more massive stars in the
early history of the galaxy is super-imposed over a standard
formation rate ($\propto$ the gas mass fraction).
 These models were proposed originally to fit the G-dwarf distribution in
 the disk (Larson 1986; Olive 1986), since it is equivalent 
to a prompt initial enrichment (PIE) (Truran \& Cameron 1971, Talbot
\& Arnett 1973). These models are found to be consistent 
with the Scalo (1986) PDMF. 
 We will explore these models to reconcile the high D destruction with
 the solar metallicity .

We shall begin by describing in section 2
our basic chemical evolution model and how
the wind  affects the evolution of D, \he3, and the heavy
elements. In section 3, we summarize the models we will be presenting.
Section 4 is devoted to discussing the ability of these models
to match the local observed properties.

\section{Chemical Evolution Models}
\subsection{Basic Equations}

A basic review of chemical evolution models can be found in Tinsley (1980).
We adopt her notation in the following discussion.  In our
models, we consider the actual stellar lifetimes thus avoiding the
instantaneous recycling approximation (IRA).  We shall be comparing the
results of our models to the observed values of the light elements in
the solar system and local ISM.  Thus we restrict our calculation
to the solar neighborhood. 

Chemical evolution models stem from tracing the evolution of the gas mass 
in the disk of the galaxy. In this study, we will consider only
the effects of gaseous flows out of the disk.
Thus the equation for the gas mass is given by
\begin{equation}
\frac{dM_G}{dt}=-\psi (t)+e(t)-o(t)
\label{1}
\end{equation}
\noindent In this equation, $\psi (t)$ is the rate at which gas is being 
used up by star
formation.  The rate of outflow of gas from the disk, $o(t)$, will be
determined by our galactic wind model and will be
discussed in detail in section 2.2 below. The rate at which gas is
returned to the ISM by mass loss or stellar deaths either in supernova events 
or in
planetary nebulae, $e(t)$, is given by 
\begin{equation}
e(t)=\int_{m_1(t)}^{m_{up}}(m-m_R)\varphi (m,t)\psi (t-\tau (m))dm, 
\end{equation}
\noindent where $m_1(t)$ is the mass of a star which dies at age $t$, and 
$m_{up}$ is the upper mass limit of stars which will supernova. $\tau
(m)$ is the main sequence lifetime of a star of mass $m$.  
Stellar lifetimes are adopted from Scalo (1986). 
$m_R$ is the mass of the
left over remnant (white dwarf, neutron star or black hole). In our
calculations, $m_R$ is taken to be (Iben \& Tutukov, 1984), 
\begin{eqnarray}
m_R & = & 0.11m + 0.45  M_{\odot} \quad m \leq 6.8  M_{\odot}, \\ 
m_R & = & 1.5  M_{\odot} \quad m > 6.8  M_{\odot}.
\end{eqnarray}
The IMF, $\varphi (m,t)$, is allowed to be both a function of mass and
time and is normalized such that 
\begin{equation}
\int_{m_{low^{}}}^{m_{end}}m\varphi (m,t)dm=1.
\end{equation}
In this equation, $m_{end}$ is the upper mass limit of stars which can
form.
In principle, there could be a distinction between $m_{end}$ and
$m_{up}$ in these equations.  It has been suggested (Larson 1986, Olive \etal
1987) that one way to avoid the overproduction of $^{16}O$ in chemical
evolution models is to limit the upper mass limit size of stars which
supernova.  Stars more massive than this are assumed to collapse
entirely into black holes returning no material to the ISM. However, in bimodal
models with a rapidly decreasing SFR, $m_{up}$ is required to be low (around
16 M$_\odot$ in some cases). Because we ascribe a physical mechanism
to a galactic wind (as opposed to a low $m_{up}$), we will set $m_{up}=
 m_{end}$ and concentrate on the galactic wind.

In what follows, we 
assume bimodal star formation in most cases.  We adopt models 
similar to those in Vangioni-Flam \& Audouze (1988) and Fran\c{c}ois, 
Vangioni-Flam and Audouze (1990).

\subsection{Galactic winds and chemical evolution}

In this subsection, we outline the calculation of the abundances of the
elements including the effects of outflow.  We consider in our models
supernova-driven winds.
Our aim is to find a simple scheme to couple a wind prescription to
a galactic evolutionary model, as has been done for elliptical galaxies 
(Larson 1974, Vader 1986, David \etal 1990, Elbaz \etal 1995). 
An early  wind, indeed, could offer a
possible way to
explain large D destruction without metal overproduction. At the beginning of
the evolution, the wind removes
a significant part of the galactic gas, ejecting from the galactic disk freshly
synthesized elements. Much later, the D free ejecta of low mass stars
are diluted in relatively smaller amounts of disk material.

Note that if we use a rate of gas loss of the form $dM/dt = -\alpha \psi$, 
where $\alpha$ is constant, we recover the Hartwick (1976) solution : Z $=
y/(1+\alpha)$ in the IRA  where Z is the final 
mean metallicity and y is the yield. 
Indeed, this ``reduced yield" would moderate the metal production (see also
Wang and Silk 1993). The constant $\alpha$ depends on the escape velocity of
 the galaxy, the fraction of gas mass converted to stars (which in turn depends 
on the IMF), and the fraction of the total supernova energy which is available
for heating the gas to escape velocity and
 depends of the radiative losses of SNR. The cooling rate
is a sensitive function of the temperature, so that for $T < 5 \times 10^6$K, 
radiative losses are large and lead to thermal instabilities. The Galaxy is
presently in this regime. 

We consider a model similar to that of Vader (1986) in which 
continuous gas loss from the disk is established.  We assume that some
fraction, $\epsilon$, of the supernova energy goes 
into heating the ISM gas to the
escape velocity for the disk and subsequently leaves the system.  The
rate at which mass is lost from the system is given by
\begin{equation}
\dot{M}_W = {{2 \epsilon E_{SN} \dot{N}_{SN}}\over v_{esc}^2}, 
\end{equation}
where $\dot{N}_{SN}$ is the supernovae rate,
\beq
\dot{N}_{SN} = \int_{8}^{m_{up}}\varphi (m,t)\psi (t)dm.
\eeq
We assume that all stars of mass greater than $8$ $M_\odot$ will supernova.
 $E_{SN}$ is the amount of energy
released per supernova which we will assume to be $10^{51}$ ergs.
The escape velocity, $v_{esc}$, in the solar neighborhood is 
of the order twice the
rotational velocity i.e. $\sim 500 {\rm km s}^{-1}$ including the dark
matter halo (otherwise it would be about 300 ${\rm km s}^{-1}$). 
We shall assume that the escape velocity has not changed significantly
over the history of the galaxy in spite of the mass loss due to the
wind since dark matter dominates the
gravitational potential of the Galaxy.  

Radiative losses from the SNRs become significant if cooling begins 
before the SNRs collide and merge with one another.  We want to
estimate the fraction, $\epsilon$, of the initial supernova energy
available for heating the ISM gas to escape velocity.
We consider first the criteria for when radiative
cooling becomes important. In order for supernova remnants to overlap
before cooling, David et al (1990) have determined a critical
supernova rate $\dot{N}_{crit}$ 
which the actual supernova rate must exceed, 
\begin{equation}
\dot{N}_{crit}=0.83 {\rm kpc^{-3} yr}^{-1}\big({{n}\over {\rm{cm^{-3}}}})^{1.82}. 
\end{equation}
If the supernova rate exceeds this value then no cooling of the outer
shell has occurred before remnants collide leaving the entire energy of 
the supernova explosion
available to heat the gas. In the models we consider below, $\dot{N}_{SN} <
\dot{N}_{crit}$ and $\epsilon < 1$. To compare $\dot{N}_{SN}$ and 
$\dot{N}_{crit}$, we scale the SFR rate to 3 M$_\odot$/pc$^2$/Gyr
(Tinsley 1980) and assume a scale height of 400 pc (Rana 1991).

If cooling is important, the outer shell of the SNR will dissipate its
thermal energy very quickly
contributing no energy to drive a galactic wind.  The only source
for wind energy is then the thermal energy of the hot interior of the
remnant. Larson (1974) has estimated the residual thermal energy of a
SNR when it collides and merges with other remnants in terms of the
supernova rate and critical supernova rate,
\begin{equation}
E_r \sim 0.22 E_{SN}({{\dot{N}}\over {\dot{N}_{crit}}})^{0.32}. 
\end{equation}
Thus the fraction of energy available to drive the galactic wind,
$\epsilon$, is given by
\begin{equation}
\epsilon = 0.22({{\dot{N}}\over {\dot{N}_{crit}}})^{0.32}. 
\label{ep}
\end{equation}

When the disk of the Milky Way was young, the gas mass fraction was larger
by a factor of about 10, the scale height of the gas was probably much higher
than it is now (say 1 kpc), and the radius of the disk was possibly
larger.  We estimate then that the initial gas density was perhaps 5
to 10 times 
greater than it is today. To determine the 
number density of gas at previous times, we scale the number
density with the evolution of the gas mass fraction.  We choose a
density of the form 
\begin{equation}
n = \big({{\sigma}\over {0.1}}\big)n_{today},
\end{equation}
where the number density today in the solar neighborhood is 0.5
cm$^{-3}$.

Since we shall be considering bimodal models where the supernova rate is
much higher in the past than it is today, the possibility exists that
more gas than can be swept up by a SNR can be heated to the escape
velocity. Thus we need to impose this constraint on our mass loss
mechanism.  The shell radius at which cooling becomes significant is
given by (Larson 1974),
\begin{equation}
R_c = 27 n^{{-7\over {17}}},
\end{equation}
Thus the amount of mass that can be swept up by a SNR is
\begin{equation}
M_s = {{4\over 3}}\pi \rho R_c^3 = 1859 n^{{-4\over 17}}.
\end{equation}
In reality, we should choose the radius not at which cooling becomes
important but rather the radius at which SNRs collide which is
somewhat larger. In practice,
however, we find that $M_s$ is much larger than the mass which can be
heated to escape velocity at all times.

Vader (1986) demonstrated that simple supernova-driven wind models
with a homogeneous ISM cannot reproduce the observed chemical
properties of dwarf elliptical galaxies and proposed metal enhanced
galactic wind models. 
If some fraction of supernovae ejecta which power the galactic wind
do not cool radiatively and is flushed directly out of
the galaxy, it will play no role in the chemical enrichment of the ISM. 
The remaining fraction leads to an effective yield $y(1 - \nu)$ for galactic
chemical evolution.
Thus Vader's models relied on two parameters: the efficiency $\epsilon$
which we take from Eq. (\ref{ep}), and the 
the fraction of the metals
produced in the supernova progenitors which is
blown out of the galaxy, $\nu$, which we will adjust to match the observed
metallicity in the solar neighborhood. In our most extreme cases, we take 
$\nu$ to be $\sim 0.8$, which is to be compared with $0.9$ in Vader's models.

We can now describe how we implement these mechanisms for galactic
winds. It is convenient to rewrite the mass ejected from stars 
$e(t)$ as the sum of the mass ejected in supernova events, $e_s(t)$, and 
the mass ejected by all other stars $e(t) - e_s(t)$.  Thus the
rate of mass which outflows from the disk is a combination of the mass
lost in the winds, $\dot{M}_W$, and the fraction of the supernova ejecta
which leaves the disk, $\nu e_s(t)$.   
Equation (\ref{1}) may now be rewritten as
\begin{equation}
\frac{dM_G}{dt}=-\psi (t) + e(t) -\nu e_s(t) - \dot{M}_W, 
\label{mg}
\end{equation}
and 
\beq
o(t) = \nu e_s(t) + \dot{M}_W
\label{o}
\eeq
Equation (\ref{mg}) may then be extended to solve for the time evolution of
the mass fraction of a heavy element, $X$:
\begin{equation}
\frac{d(XM_G)}{dt}=-\psi (t)X+e_X(t) -\nu e_{sX}(t) - \dot{M}_WX,
\label{xmg}
\end{equation}
where $e_X(t)$ and $e_{sX}(t)$ represent the total amount of
metals ejected by stars and the type II
supernova contribution respectively.  
Equation (\ref{xmg}) can
be further simplified to read
\begin{equation}
\frac{dX}{dt}=\frac{(\nu e_s(t)-e(t))X
-\nu e_{sX}(t)+e_X(t) }{M_G} 
\label{pre}
\end{equation}
We use this equation to calculate the abundances of the elements such as
$^{16}$O 
(For D, $e_{{\rm D}}(t) = 0$ since D is 
totally destroyed in stars.)
We have adopted the yields of Woosley \&
Weaver (1995) for
$^{16}$O. For Fe, we include contributions from both type I and II supernovae.
We take the type II yield from Woosley \& Weaver (1995).  
For type I supernovae, we take the yield
to be 0.7 M$_\odot$ (Thielemann, Nomoto, \& Yokoi 1986) for stars between
1.5 and 8 M$_\odot$ and we fix the rate of type I events to obtain a maximum
iron abundance which can be compared with the G-dwarf data below.  In order to
reproduce observational features such as the evolution of [O/Fe] verses
[Fe/H] we have built in to our models a 1 Gyr time delay corresponding to
the lifetime of the progenitors of type I supernovae
 (see e.g. Yoshii \etal 1996). (Note that the Woosley \& Weaver (1995)
iron yields are somewhat high and give a low value for  [O/Fe] vs.
[Fe/H] at very low metallicity (see eg. Matteucci \& Fran\c{c}ois 1989).)

\subsection{D and \he3}

Our goal in this subsection is to outline the evolution of D and \he3.
We begin by summarizing the observational constraints on these elements.
As discussed above, detections of D 
using quasar absorption
systems have been reported recently.  
The high values of D/H determined in these systems are ideal for demonstrating
the effects of outflow. In models where significant D destruction is required, 
the stellar processing which destroys D, produces heavy elements whose
abundances can be controlled by outflow. In these cases we take the primordial
D/H abundance to be $2 \times 10^{-4}$.
As we noted above, these measurements are still controversial and we will also
show results based on a somewhat lower value which is still consistent
with \he4 and \li7, (D/H)$_p = 
7.5  \times 10^{-5}$ a value already high from the perspective
of evolution and high enough to demonstrate the effects of outflow that we wish
to explore. For completeness, we also consider a low value for D/H = 2.5 $\times
10^{-5}$ as inferred in some observations of quasar absorption systems.

The present day D
abundance has recently been determined by Linsky \etal (1993, 1995) using the
HST.  They determined an ISM D/H abundance of
\beq
{\rm (D/H)_{ISM}} = 1.60 \pm 0.09 ^{+0.05}_{-0.10}
\times 10^{-5}
\eeq
 The present day \he3 abundance has been determined in a number
of galactic HII regions (Balser \etal 1993).  These values range from $1.1
- 4.5 \times 10^{-5}$. A recent measurement of \he3/H by the Ulysses
 satellite by Gloeckler and Geiss (1996) in low energy ions filtering from
the local interstellar medium in the solar cavity, has been performed finding
\beq
 {\rm (\he3/H)_{ISM}} =2.2 ^{+0.7}_{-0.6} \pm 0.2 \times 10^{-5}
\eeq
 and is complementary to the local D/H measurement.
 We will take this measurement as a reference for the present and local galactic 
environment.  

The presolar D and \he3 abundances were recently reviewed in Geiss
(1993) and discussed in Scully \etal (1996).  Presolar D is not directly
measured.  Instead, a comparison is made between the \he3 abundance measured in
the low temperature components of carbonaceous chondrite meteorites (which are
in good agreement with the measured \he3/H ratio in the lunar soil and solar
wind) and the abundance measured
in high temperature components these meteorites. 
The latter \he3 measured in the carbonaceous chondrites represents a true
presolar abundance for this element. Since it is assumed that all of the D
present in the gas from which a star is formed is converted
to \he3 in the pre-main sequence phase, the other
measurements really represent D + \he3 present before the formation
of the solar system.  Our adopted presolar values of D, \he3,
and D + \he3 are (Scully \etal 1996):
\begin{equation}
[(D + ^3He)/H]_{\odot} = (4.1 \pm 0.6 \pm 1.4) \times 10^{-5},
\end{equation}
\begin{equation}
(^3He/H)_{\odot} = (1.5 \pm 0.2 \pm 0.3) \times 10^{-5},
\label{he3}
\end{equation}
\begin{equation}
(D/H)_{\odot} = (2.6 \pm 0.6 \pm 1.4) \times 10^{-5}.
\end{equation}
However, recent measurements of surface  abundances on Jupiter
show a somewhat higher value for D/H,  D/H = $5 \pm 2 \times 10^{-5}$ 
(Niemann \etal 1996) and D/H = $5.9 \pm 1.4 \times 10^{-5}$
(Ben Jaffel \etal 1996). If these values are confirmed and if fractionation
does not significantly alter the D/H ratio (as it was suspected to for 
previous measurements involving CH$_3$D), they should be taken 
into account in galactic chemical evolution models.  As it stands, these values
are marginally consistent with the {\it inferred} meteoritic values.  In a 
chemical evolution model, they can be made consistent with the high
QSO absorber D/H values, but not the low ones which would then require
the net production of D/H.

The evolution of D in the ISM can be determined by
extending equation (\ref{xmg}), 
\begin{equation}
\frac{d(M_GD)}{dt}=-\psi (t)D - \dot{M}_W D.
\end{equation}
By substituting equation (\ref{mg}) into the above equation and with further
simplification, the D evolution is governed by,
\begin{equation}
\frac{dD}{dt}=\frac{(\nu e_s(t) - e(t))D}{M_G} .
\end{equation}

More complicated than the history of deuterium is that of \he3. In more
massive stars ($M < 5-8 M_{\odot }$) , \he3 is destroyed on average
(although not completely). In lower mass stars (1 $M_{\odot }\leq M\leq $ 2$%
M_{\odot }$), it is likely that \he3 is produced. Iben and Truran (1978)
estimate the \he3 in the surface layers of stars $<8 M_{\odot}$ to be 
\begin{equation}
(\frac{^3He}H)=1.8\times 10^{-4}(\frac{M_{\odot }}M)^2+0.7[\frac{(D+^3He)}%
H]_i^{} 
\label{it}
\end{equation}
\noindent where the bracketed term is the contribution from the D in the
pre-main sequence phase. Similarly, significant production of \he3 was found by
Vassiladis \& Wood (1993) and by Weiss, Wagenhuber, and Denissenkov (1995).
There is in addition, some observational support for the production of \he3 in low
mass stars as evidenced by the high \he3/H ratios measured in planetary nebulae
(Rood, Bania \& Wilson 1992, Rood \etal 1994).
 
There has been a considerable amount of discussion recently concerning the
\he3 production in low mass stars. It has been argued 
(Charbonel 1994, 1995,  Hogan 1995, Wasserburg, Boothroyd, \& Sackman, I.-J.
1995, Weiss \etal 1995, Boothroyd \& Sackman 1995, Boothroyd \& Malaney 1995)
that perhaps low mass stars in the range 1--2 M$_\odot$ are in fact net
destroyers of \he3. As it appears unlikely that chemical evolution alone can
resolve the problem concerning the overproduction of \he3 (Scully \etal 1996),
these latter yields for low mass stars should be taken seriously.
For larger mass stars, we adopt the \he3 survival fraction given by
Dearborn, Schramm, \& Steigman (1986) which is consistent with the 
recent results of Woosley \& Weaver (1995).
The \he3 evolution may now be determined by following
the prescription of equation (\ref{pre}). We will consider an alternative
approach to \he3 in \S 5.

\section{Models}

We will consider a model with a bimodal star formation rate which includes the
galactic wind described in section 2.2.  The specific model has been
chosen to satisfy observational constraints on D for the solar epoch
and present day (we will consider different
primordial values of D/H which thus affects the parameters of the model).
  The parameter $\nu$ is adjusted to reproduce the
observed solar $^{16}O$ abundance.  We also consider only models which
can reproduce a gas fraction today in the range of $\sigma = 0.05 - 0.20$
which roughly corresponds to the observed gas fraction (Rana 1991). 

The first case we consider has a primordial D/H = $2 \times 10^{-4}$ and
is clearly the most extreme case in terms of the degree of D destruction.
The bimodal star formation rate includes a
 rapidly decreasing exponential SFR associated with
an IMF favoring more massive stars which is superimposed on top of
a SFR proportional to the gas mass associated with a more normal IMF.
This model which we will refer to as model Ia is similar 
to model IV of Vangioni-Flam and Audouze (1988).  The exponential
component includes a SFR $\psi_2(t) = 0.29e^{-t/2}$ with an IMF $\phi(m)
\propto m^{-2.7}$ in the mass range of $2 \le m/M_\odot < 100$.  The
more normal component has a SFR $\psi_1(t) = 0.29M_G$ with an IMF
$\phi(m)\propto m^{-2.7}$ in the mass range of $0.4 \le m/M_\odot <
100$.  We find that about 81\% of the supernova ejecta or $\nu = 0.81$
is necessary to reproduce the observed solar oxygen abundance.
This is to be compared with values as large as 0.9 found by
Vader (1986). 

As we indicated earlier, the type I to type II supernova rate is 
determined by the produced iron abundance in the model. In fact, the
iron abundance depends on three quantities, the type I iron yield, the
fraction of stars becoming type I supernovae, and the fraction of
type I ejecta expelled in winds.  The latter need not be equal to 
$\nu$ which corresponds to the type II fraction ejected. Our models
rely only on the product of all three of these parameters. 
In model Ia,
the present type I to type II SN ratio 
(when we refer to this ratio, we will always be referring to its present-day
value) is 
low (2\%) if we assume that no type I ejecta 
is expelled, whereas
the ratio is 12\% if we assume that winds carry the same fraction of 
type I ejecta as for type II. The fact that this ratio is low
also reflects the high Fe yields for type II SNe from Woosley \& Weaver (1995).
  Note that the type I to type II SN
ratio should be in the range of 10 -- 20 \% to be consistent with
observations (Tammann 1994).

When we consider the G-dwarf distribution, 
we will compare the above model to one in which
the massive star component is followed by the more normal
component sequentially.  For this model we consider an exponential
SFR $\psi_2(t) = 0.19e^{-t/1}$ for $t \le 1$ Gyrs 
with the same massive IMF as model Ia and a SFR $\psi_1(t) = .73e^{-t/2.5}$
for $t > 1$ Gyrs with the same normal mode IMF as model Ia. In this case,
we find that a
value of $\nu = .68$ is necessary to reproduce the observed solar 
oxygen abundance. The SN I to SN II ratio in this model is 4\% but can be 
as large as 13\%.
 We will refer to this case as model II.

In Scully \etal (1996) we also considered 
a model in which the IMF is a function which varies in
time (metallicity)
\begin{equation}
\varphi (m,t) \propto m^{-(1.25 + O/O_{\odot})}.
\end{equation}
The dependency of the IMF on the oxygen
abundance was chosen so that the 
oxygen abundance history predicted by the model closely matches the
observations.  This model was coupled to outflow with the expressed
purpose of reducing the the abundance of \he3.  While it alleviated
the problem somewhat, \he3 was nevertheless overproduced resulting in the 
conclusion that chemical evolution alone could not suffice in 
resolving the \he3 problem. The wind model here was not ``designed"
to fix \he3 and in fact because of large portions of ISM are blown 
out the \he3 abundance actually increases when outflow is included.
Overall, this type of model (with the time varying IMF) does not give
qualitatively different results.

We also consider models which require far less destruction of D/H.  For a
primordial value D/H = $7.5 \times 10^{-5}$, we have in model Ib,
a SFR, $\psi_1(t) = 0.28M_G$ and $\nu = 0.55$. 
In this case the SN I to SN II ratio is only
3 -- 7 \%. This is not a bimodal model.
The same IMF as in the normal mode
in model Ia was used. The associated \he4 and \li7 abundances predicted by
BBN are consistent with their observationally determined values.
Finally for completeness, we considered a model
Ic, in which the primordial D/H was set to $2.5 \times 10^{-5}$.
In this case, we ran a model with
a simple constant SFR, $\psi = 0.07$.  This value of $\psi$ is chosen
to obtain a suitable evolution of D/H.  Because heavy element {\em pro}duction
rather than over-production is a problem, $\nu = 0$. In this case, the lower
mass limit in the IMF was lowered to 0.2 M$_\odot$ and the SN I to SN II
ratio is 1\%. Once again, this low ratio reflects the high type II iron yields.
Recall that this ratio could be adjusted upward by lowering the assumed 
type I yield.
 This value of D/H was chosen to correspond to 
the low D/H observed in certain quasar absorption systems (Tytler, 
Fan, \& Burles 1996, Burles \& Tytler 1996).  However, it 
should be noted that the predicted \he4 and \li7 abundances are not in 
agreement with the observations.

\section{Observational Constraints and Results} 

\subsection{Element Abundances} 

In figure 1, we show the evolution of D/H as a function of time for
model Ia along with the evolution of \he3/H.  
This model is capable of
adequately explaining the evolution of D/H even with the high
primordial abundance of D/H = $2.0 \times 10^{-4}$.  Figure 2
illustrates the corresponding oxygen evolution for model Ia which we
use as a tracer for heavy element production. In both figures 1 and 2,
the dotted line shows the evolution of a closed box model
(no outflow) with the SFR and IMF of model Ia, the solid line in contrast 
shows the effect of the galactic winds as described in section 2.2.
 With the inclusion of a
metal enriched galactic wind, we are able to find models which can
destroy a sufficient amount of D without overproducing metals. The 
present gas mass
fraction in this model is 0.07 (it turns out to be 
between 0.07 and 0.10 in all of the models considered). 
Figures 1 and 2 also show the resulting
evolution of model II. 

With the D destruction necessary for our models to reproduce the
observed solar
and present-day D abundances, it is not surprising that if
we take the Iben \& Truran (1978) yields for \he3, we find a 
a solar \he3 abundance of about $\sim 10$ times higher than is
observed.  Enriched outflow makes the problem worse with regard to
\he3. A comparison of model Ia without outflow is also illustrated on figure 1.
\he3 increases more rapidly with outflow than without.  This is
because we are including in our wind a fraction of the supernova
ejecta which is depleted in \he3 relative to the ISM abundance.  There
is therefore less \he3 poor gas available to dilute the ISM.  We will
discuss ways of resolving this problem in \S 5 below.   

In figures 3 and 4, we compare the results of the three cases with differing
initial D/H.  In model Ib, which results in a very similar evolution to
the model considered in Olive \etal (1995), \he3 is still greatly overproduced.
In Ic, where the D/H abundance starts out very low, \he3 is consistent 
within the errors if systematics are included.  
The abundance of \he3 at the solar epoch is 
\he3/H $= 2.0 \times 10^{-5}$ which is to be compared with the solar value
in Eq.(\ref{he3}). We point out however that this is not an entirely acceptable
model. First, we still destroy a bit too much deuterium. 
This could be fixed by including some primordial infall.
However, infall would aggravate both the slightly low metallicity (we do not
quite achieve solar metallicity at the solar epoch) and the somewhat high 
present day gas mass fraction (28\%).  Primordial, D-rich 
infall would further lower the metallicity and increase the gas mass fraction.

A considerable amount of enriched material is expelled from the galaxy
in this type of model (Ia). This could contribute to the enrichment of the 
extra-galactic gas with
heavy elements. The same process is strengthened in clusters of
galaxies where ellipticals are dominant and provide most of the 
metals to the intracluster gas (Elbaz \etal 1995) as observed by X-ray
satellites (Mushotzky \etal 1996, Loewenstein \& Mushotzky 1996).  An overall
metallicity of about 0.1 -- 0.2 $Z_\odot$ results in the gas surrounding the
galaxy from the enriched outflow of model Ia.  
Thus we can conclude that these models do not
expel an unreasonable amount of metals from the Galaxy.   

The amount of outflow, $o(t)$, as defined in equation (\ref{o}),
produced by supernovae driven winds is shown
as a function of time in figure 5 for model Ia, b, c, and II. 
As it is directly tied to the
supernova rate, it is a sharply decreasing function of time.
In figure 6, the effect of outflow on the mass of Galaxy is shown.
Despite the large SFR early on and the degree of D destruction in the model, 
the mass of the Galaxy only changes by about 30\% in Ia, by 8 \% in Ib
and less than 3\% in Ic. In model II, the mass changes by about 14\%.

\subsection{The G Dwarf Distribution}

The distribution of G-dwarf stars as a function of metallicity serves as 
a constraint for chemical evolution models. G-dwarfs have sufficiently
long lifetimes so that most of them which formed early on in the galaxy should
be present today.  The problem as normally stated is that
there is a lack of metal-poor stars observed relative to the predictions of
simple closed-box chemical evolution models. A number of solutions
including a prompt initial enrichment (PIE) (Truran and Cameron 1971),
inflow of processed material (Ostriker and Thuan 1975), and accretion
of unprocessed material (Larson 1972) have been proposed. Any
realistic model 
of the 
chemical evolution in the solar neighborhood 
should account for the lack of metal 
poor dwarf stars.
 
The rate of formation of G-dwarf stars is given by
\begin{equation}
{dN\over dt} = \int_{m_l}^{m_h}\varphi (m)\psi (t)dm, 
\end{equation}
where $m_l = 0.8$M$_\odot$ and $m_h = 1.1$M$_\odot$ are the 
lower and upper mass limits of G-dwarf stars formed at time 
$t$.  Bazan and Mathews (1990) point out that some
of the G-dwarfs which formed early on would have had short enough lifetimes 
that they would not be present today.  In order to account for this, 
the number 
of G-dwarf stars which have formed must further be multiplied by 
the fraction of those which survive given by
\begin{equation}
f(t_g -t) = \int_{m_l}^{m(T_G-t,Z(t))}\varphi (m)dm / 
\int_{m_l}^{m_h}\varphi(m)dm,
\end{equation}
where $m(T_G-t,Z(t))$ is the largest star of metallicity $Z(t)$ which is 
left after a time $T_G -t$.  

Before we discuss the distribution of G-dwarfs for the chemical evolution
models we consider, we first note the effect of (dropping) the IRA.  
In figures 7-9, we compare a simple closed box model with and without
the IRA. In figure 7, we show the cumulative number of G-dwarfs as a function of 
metallicity taken here to be [Fe/H] (scaled by the Fe abundance
at $t = 14$ Gyr). The data represented by
points are taken from Pagel (1988).  The  curves show 
the G-dwarf totals using a simple closed-box model with $\psi
= 0.25\sigma = 0.25 M_G$ 
and an IMF $\varphi(m) \sim m^{-2.7}$ in the range $0.4 \le
m/M_\odot \le 100$ with (dashed) and without (solid) the IRA.
There is a significant difference here which can be explained.
In figure 9, we show the age metallicity relation for the two cases
described above.  In the IRA, the metallicity increases linearly with time,
and as such a
 given metallicity bin corresponds to a relatively small time
bin, so that at late times the product of a diminishing SFR and a small
integration time produces few G-dwarfs.  When the total number of G-dwarfs
is normalized to the observed total (132 in this case), we see an excess
at low metallicity.  This is the classic G-dwarf problem.  When the IRA
is dropped, the Fe/H is no longer linear in time and the age-metallicity
curve flattens toward higher metallicity corresponding to late times.
This is due to the non-negligible lifetimes of lower mass stars,
whose ejecta at late times dilutes the metallicity.  
As a result, the high metallicity bins correspond to 
significantly large time bins, and many more G-dwarfs are computed 
to be produced at higher metallicity.  Because we normalize to the same
total number of dwarfs, there are far fewer formed at early times or 
low metallicity.  Thus the problem is shifted to a deficiency of G-dwarfs at
intermediate metallicities (around [Fe/H] $\approx -0.5$) and an excess
at near solar metallicity. This problem is evidenced in the models of Timmes,
Woosley \& Weaver (1995) where the IRA is not employed.

Pagel (1988) advocates that in addition to the cumulative number of G-dwarfs
it is useful to 
 plot the change in the number of dwarf stars as a function of 
metallicity. Hereafter, we will refer to this as the differential
G-dwarf problem. In figure 8, we show the differential distribution of 
G-dwarfs vs [Fe/H] for the simple closed box model with and without the IRA.
Typical errors in the data are $\sqrt{\Delta N}$ for the number of
dwarfs in each bin and 0.1 dex in [Fe/H].
Again we see clearly the shift in the excess number of dwarfs to 
higher metallicity. In addition, notice that the turnover at higher
metallicities is absent when the IRA is dropped.
 We conclude that any attempt to resolve the 
G-dwarf problem can not be based on the IRA.

In figure 10, we show the G-dwarf predictions 
from the bimodal models Ia, and II and model Ic. 
The curve for Model Ib looks very similar to that of Ic and is not shown.
 It is demonstrated that the models
which include more massive star production at earlier times
predict an abundance of metal poor 
stars which is more consistent with the observations.  
This is a result of the rapid
rise of the metallicity of stars due to the rapid production of heavy elements
associated with the massive star production. However, with the exception of 
model II (which was constructed to produce a G-dwarf distribution 
which more closely matches the observations), these models show a
deficiency of dwarf stars at metallicities a factor of 2-3 below solar.
This is the same problem witnessed in the closed box models when the
IRA was dropped

Figure 11a shows the differential G-dwarf distribution for 
models Ia, c and figure 11b shows the result for model II.  
For all of the models considered, we find that there is no 
(or very little) excess of 
G-dwarfs at low metallicity.  In contrast, we see clearly from the differential
distribution that 
with the exception of model II, models Ia, and c show a deficit
of dwarfs at metallicities somewhat below solar and an excess at
metallicities just above solar (model Ib is similar). Thus contrary
to what is often claimed, the G-dwarf distribution has very little
to do with the total amount of astration of deuterium.
Large deuterium destruction factors can not be excluded
on this basis. It should be noted that 
discrepancies in the highest of
the metallicity bins may be artificial.  The data (Pagel 1988) show 
7 out of 132 stars in the final bin at [Fe/H] between 0.1 and 0.2.
We have tuned these models by adjusting 
the rate of type I supernovae and produced a turn over
in the final metallicity bin. As in the case of the closed box models, 
these models result in an oxygen abundance as a function
of time which flattens out after a sharp rise (figure 6).  This makes for a bin
size for metallicities of around solar which cover several billions of years.
Thus even though the SFR is lower at these times, it is more than compensated 
for by the longer time period for the formation of G-dwarfs.  Thus the rise in
the number of G-dwarfs towards higher metallicities. Aside from the turn over
at higher than solar metallicities, these distributions are qualitatively 
similar to the the distribution shown in Timmes, Woosley \& Weaver (1995).
Model II does not exhibit this problem due to the rapidly
decreasing SFR (exponential with a time constant of 2.5 Gyr).  
As demonstrated in Fran\c{c}ois \etal (1990), the sequential model
is best suited for obtaining an acceptable G-dwarf distribution as the massive
mode in this case acts as a PIE. In this case, during the first 1 Gyr,
there is significant metal production, bringing [Fe/H] close to -1.0 without
the production of any dwarf stars. The ``normal" 
mode is then shifted towards higher metallicity.

We have also checked the white dwarf production rates and accumulated
white dwarf surface density for model II.  If we define the white dwarf birth
rate by 
\beq
B_w(t) = \int_{m_1(t)}^{m_w} \phi(m) \psi(t-\tau(m)) dm
\eeq
where $m_w$ is the upper mass limit for the formation of a white dwarf,
taken here to be 8 M$_\odot$. The present white dwarf birth rate in model
II is 1.4 $\times 10^{-3}$ pc$^{-3}$ Gyr$^{-1}$, consistent with 
the rate found by Weidemann 
(1977).  The total integrated surface density of white dwarfs is
9 M$_\odot$ pc$^{-2}$ and is somewhat smaller than the 
white dwarf density found in Larson's (1986) bimodal models.
Finally, the white dwarf luminosity function $n = B_w(t-t_c)
\Delta t_c/2\langle z \rangle$, where $t_c$ is the cooling time (taken from 
Iben and Tutukov (1984)) $\Delta t_c$ is the time to cool from $M_{\rm bol}
-.5$ to $M_{\rm bol} + .5$, and $\langle z \rangle$ is the scale height.
In model II, the luminosity function is somewhat high at very low luminosity,
$n$ appraoches 10$^{-2}$ pc${^-3}$ mag$^{-1}$ at $M_{\rm bol} = 17$
though it is lower than that predicted in Larson's (1986) model (Olive 1986).

Finally, in figure 12, we show the age metallicity relation for the 
models discussed above and compare it to the data of Edvardsson \etal (1993).
As one can see, all of our models are well within the established scatter
in the observations.  Indeed, from the data on the [Fe/H] vs age, 
it is clear that the Galactic metallicity remains rather flat
(at about solar metallicity) for most of the history of the 
Galaxy (the last 10-12 Gyr). The large integration time for the production
of dwarf stars at solar metallicity, makes it difficult to reconcile
the large number of G-dwarfs observed around [Fe/H] = -0.5, where
the integration time is relatively short.  This can only be done 
in models in which the star formation rate is a steeply decreasing
function of time. We emphasize again that this problem is very
different from what is commonly referred to as the G-dwarf problem.
Dropping the IRA, is nearly sufficient for resolving the problem
of an excess number of dwarfs at low metallicity.

\subsection{The PDMF}

The only direct observational constraint on choosing an IMF is the 
present-day mass function (PDMF).  The PDMF refers to those stars on
the main sequence which are still observable today.  The PDMF has been
estimated from the luminosity function by Scalo (1986).  We shall
adopt his results for comparison with our models.

For stars which are born and have
lifetimes greater than the age of the galaxy or roughly masses $<
0.9M_\odot$, the IMF and PDMF are directly comparable.  All of these
stars with lifetimes, $T_m$, are still on the main sequence so the
PDMF, $\varphi_{MS}$, for these stars is just the total number born given by 
\begin{equation}
\varphi_{MS} =  \int_{0}^{T_G}\varphi (m,t)\psi (t)dt  \qquad ~~~~ T_m \geq
T_G,
\end{equation}
where $T_G$ is the age of the disk and $T_m$ is the main sequence
lifetime of a star of mass $m$.
For stars more
massive than this, the PDMF represents the number of stars of a given
mass
which have
not yet evolved off of the main sequence.  Thus only those stars born
in the last $T_m$ years will be on the main sequence. Therefore, the
PDMF for these stars is given by
\begin{equation}
\varphi_{MS} = \int_{T_G-T_m}^{T_G}\varphi (m,t)\psi (t)dt   \qquad T_m <
T_G.
\end{equation}

We have chosen a simple power law IMF for model Ia with
$\varphi \propto m^{-2.7}$ consistent with the Scalo PDMF with some
simple assumptions about the SFR (Olive \etal 1987).  We do include a
massive star component in these models but in each case the massive
component steeply decreases very early on.  We therefore expect that
the increased production of massive stars early in the galactic
history will have no effect on the PDMF predicted by these models
since the larger number of massive stars created early on would have
long since died out.
Figure 13 shows the PDMF plotted along with the data from
Scalo (1986).  The PDMF predicted from the models is in good agreement
with the Scalo data for stars $> 0.9 M_\odot$.  The low mass end of the PDMF
could be fit by a multislope IMF, which flattens out at low masses. This 
would have no effect on the issues considered here, as the only stars which
affect the chemical evolution of the Galaxy are those with masses greater than
about 0.9 M$_\odot$.  Below this limit, stars are so long lived that they have
not ejected their processed material.  Though flattening the IMF at low masses
would affect the amount of gas trapped in stars, this can be compensated
by adjusting $m_{low}$.
A similar result was found in Scully \etal (1996) for models with
a metallicity dependent IMF. In this case, even though a larger
number of massive stars is created early on, the IMF very quickly
steepens to a more normal IMF ($\varphi \sim m^{-2.6}$) so no evidence
of the massive stars created early on shows up in the PDMF which
results from this type of model. As these are the most extreme
cases, we do not show the PDMF for the models with smaller primordial
D/H.

\section{Implications for \he3} 

It is clear from the results above and from the work of Scully \etal
1996, that the problems concerning \he3 can not be fully resolved
by galactic chemical evolution alone.  As is well known, the root
of the problem concerning \he3 is the production of \he3 in low
mass stars as given for example in Eq.(\ref{it}). Models in which \he3 is
at least partially destroyed in low mass stars fare much better 
( Vangioni-Flam \etal 1994, 
Vangioni-Flam \& Cass\'{e} 1995, Olive \etal 1995, and Scully \etal 1996)
and are 
now being put on a firm astrophysical basis by stellar modelists themselves 
who invoke an extra mixing mechanism due to diffusion below the 
convective envelope, possibly driven by rotation (Charbonnel 1995, 1996, 
Hogan 1995, Wasserburg \etal 1995,  Boothroyd \& Sackman 1996, Weiss, 
Wagenhuber \& Denissov, 1996). This mechanism seems, at the same 
time, to explain the high $^{13}$C abundance in globular cluster red giants (see
 e.g.   Boothroyd \& Sackman (1996), for an extensive discussion). However,
the  important measurement of Gloeckler \& Geiss (1996) leads us to reanalyze 
the situation.

The new measurement indicates that the sum (D + \he3)/H seems to be approximately 
constant since the birth of the sun, varying from 4.1 $\pm 0.6 \pm 1.4
 \times 10^{-5}$
 to 3.8 $\pm 0.7 \pm 0.2 \times 10^{-5}$ (Turner \etal 1996).
(If the measurements of D/H in Jupiter turn out to be representative of the
protosolar abundance, then (D+\he3)/H may actually be required to decrease
over the last 4.6 Gyr.)
 If $T_g$ is the galactic age in Gyr, the late \he3 behavior is 
governed by stars of lifetimes longer than $T_g - 4.6$, 
which corresponds to a very narrow mass range: $M(T_g)$ to $M(T_g - 4.6)$.
For $T_g = 14$ Gyr, this mass range corresponds to 0.9 -- 1.0 M\sun.
 In addition, the relative constancy of the \he3/H abundance from the 
birth of the Galaxy up to the birth of Sun would implies that above a critical
mass,  $M_c$, $(M_c > M(T_g-4.6))$, \he3 is significantly destroyed. However,
as noted above, the presence of a high observed \he3/H abundance in planetary
nebulae (Rood \etal 1992, 1995) indicates that not all stars can efficiently
destroy \he3.

Given an age of the Galaxy, $T_g$, once the empirical D history is 
fit by adjusting the essential parameters of galactic evolution, the model 
allows a clear empirical determination of the mass 
of stars which contribute to the 
\he3 enrichment from $T_g-4.6$ Gyr until now, and also the degree of 
production and destruction of this isotope in different mass ranges.
Stellar lifetimes are key ingredients of the model, especially 
those of low mass stars that arrive to maturity and shed their processed 
material since the birth of Solar System.

Another influential parameter, traditionally called $g_3$, is defined 
as the \he3/H ratio in the ejected material divided by the initial one i.e. 
(D+\he3)/H at star formation. 
We have taken $g_3$ to be given by Eq.(\ref{it}) for stars with
masses less than M$_c$, ie. these stars are net producers
of \he3, while for stars more massive M$_c <$ M $< 8$M$_\odot$,
we have taken $g_3 = 0.1$ and for more massive stars
 we take $g_3$ from Dearborn,
Schramm, \& Steigman (1982). We find that
in all of the cases studied, the transition mass is always 
M$_c \simeq $ 1.0 M$_\odot$, thus the \he3 producers lie in a very narrow range 
as argued above.
In figure 14, we show the evolution of D/H and \he3/H for model II,
assuming the above model for \he3.  In this case the mass cut is at 0.96
M\sun.

{}From this analysis, a typical case emerges corresponding to $M_c \sim 
1.0$ M$_\odot$ corresponding to our reference model. Below this mass, \he3 is 
produced and above it is destroyed. Thus it is expected that a few stars, in a 
limited mass range, do produce \he3 as observed in certain planetary 
nebulae (Rood et al 1995).  Since the favorable mass range is so narrow (0.9 to 
1.0 compared to  that of all PN progenitors (0.9 to 8 M$_\odot$), it is not
surprising  that such \he3 rich objects are rare.  Indeed, for a standard IMF we
get about 17\% (about 11\% in the case depicted in figure 14), 
which seems reasonable.

Concerning the nucleosynthesis of \he3, according to our 
calculation, the mass of the \he3-rich PN progenitors, which should have 
been born in the very early Galaxy, is close to 1 M$_\odot$. Unfortunately
at present we can only consider such a model as empirical as it has no other
real theoretical foundation. It is worth noting however,
that the mass range of interest  covers the region where
related phenomena take place: 1) a central convective core forms which could
 modify the subsequent evolution
(see eg. Schaller \etal 1992);  2) the energy production
rate of the CNO cycle begins to overcome that of the p-p chain 
(see eg. Arnett 1996). The model above is at variance with
models which invoke additional mixing
in order to explain the enhanced $^{13}$C abundance in red giants,
and require that the bulk of the \he3 destruction takes place
in stars with $M < 2 M_\odot$. 
There are however several problems with the models which destroy
\he3 only at the low end of mass function.  First, there is generally not enough
destruction of \he3 to account for the low solar \he3/H abundance (Scully \etal
1996). Though they aid significantly, \he3 is still overproduced by a factor of
about 2 at the solar epoch. In addition, if the progenitors of the \he3 rich 
planetary nebulae are relatively massive ($M > 2 M_\odot$), then it is difficult
within standard stellar models to explain the necessary enhancement of \he3
 (\he3/H $\sim 10^{-3}$) in these systems (cf. Eq.(\ref{it})).
The model of Boothroyd and Malaney (1996)
on the other hand, does provide enough \he3 destruction, yet are even
more problematic as far as the planetary nebulae data are concerned
 (Olive \etal 1996).

\section{Conclusion}    

The high D/H ratio inferred from the absorbing clouds along the line of 
sight of certain remote quasars led us to go beyond  standard galactic
evolutionary schemes which can not account for a deuterium destruction 
factor greater than a few. To this aim, we
have relaxed the closed box hypothesis and coupled a a simple galactic
wind model to the galactic evolutionary one.
The wind is driven by the numerous core collapse supernovae that 
are assumed to explode in the early galaxy, due to a somewhat 
enhanced massive star formation rate early on and rapidly diminishes 
at later times.  The early generation of massive stars has
three beneficial effects: 1) they quickly eject their D-free material and
induce a lowering of the D/H ratio; 2) they induce a galactic outflow as 
long as their explosion rate remains sufficiently high.  This wind
moderates the rise in metallicity; 3) they rapidly enrich the interstellar
medium to such a level that the G-dwarf problem is alleviated, if not solved.
With regard to the G-dwarf problem, we have emphasized that the standard 
excess of dwarfs at very low metallicities is largely due to the IRA. 
When the IRA is dropped, the problem is shifted to a {\em deficiency} of 
dwarfs at [Fe/H] $\sim -0.5$ and an excess at solar [Fe/H] and above.
We have been able to match all of the observational constraints
considered, and in particular a D destruction factor of about 10, while
maintaining a solar metallicity.

The evolution of \he3 is more problematic and remains the least understood
of the light element isotopes. We believe that the difficulty 
in the abundance patterns of \he3 resides with stellar evolution
rather than with galactic chemical evolution or cosmology.
Using the recent observations of Gloeckler \& Geiss (1996), we have shown
empirically that stars above 1.0 M$_\odot$ destroy \he3 thoroughly and 
produce it at lower masses in qualitative agreement with classical 
stellar evolution theory and the observations of \he3 in planetary nebulae.
          
\section{Acknowledgments}   
We would like to thank K. Nomoto, D.N. Schramm, G. Steigman, and J.W. Truran
for helpful discussions.  We would especially like to thank Bernard Pagel
for his helpful comments on our manuscript.
This work was supported in part by PICS n$^\circ$114, 
``Origin 
and evolution of the light elements," CNRS. 
This work  was  also supported 
in part by  DOE grant DE-FG02-94ER-40823.

\newpage
\beginapjbib

\bibitem Arnett, D. 1996, {\it Supernovae and Nucleosynthesis}
 (Princeton University Press, Princeton)

\bibitem Balser, D.S., Bania, T.M., Brockway, C.J.,
Rood, R.T., \& Wilson, T.L. 1994, ApJ, 430, 667

\bibitem Bazan, G. \& Mathews, G.J. 1990, ApJ, 354, 644

\bibitem Ben Jaffel, L. \etal 1996, Science, submitted

\bibitem Boesgaard, A. 1996, to appear in {\it Formation of the Galactic Halo:
Inside and Out}, eds. H. Morrison \& A. Sarajedini (Astronomical Society
of the Pacific Conference Proceedings)

\bibitem Boothroyd, A.I. \& Malaney, R.A. 1995, astro-ph/9512133

\bibitem Boothroyd, A.I. \& Sackman, I.-J. 1995, astro-ph/9512121

\bibitem Burles, S. \& Tytler, D. 1996, astro-ph/9603070

\bibitem Carswell, R.F., Rauch, M., Weymann, R.J., Cooke, A.J. \&
Webb, J.K. 1994, MNRAS, 268, L1

\bibitem  Carswell, R.F., \etal. 1996 MNRAS, 278, 518

%\bibitem Cass\'{e}, M. \& Vangioni-Flam, E. 1994,
%talk presented at the ESO/EIPC Workshop on
%the Light Element Abundances

\bibitem Charbonnel,C. 1994, A \& A, 282, 811

\bibitem Charbonnel,C. 1995, ApJ, 453, L41

\bibitem Charlton, J. \& Salpeter, E.E. 1989, 

\bibitem Dar, A. 1995, ApJ, 449, 550

\bibitem David, L.P., Forman, W., \& Jones, C. 1990, ApJ, 359, 29

\bibitem  Dearborn, D. S. P., Schramm, D.,
\& Steigman, G. 1986, ApJ, 302, 35

\bibitem Dearborn, D., Steigman, G. \& Tosi, M. 1996, ApJ, 465, in press.

%\bibitem Delbourgo-Salvador, P., Gry, C.,
%Malinie, G., \& Audouze, J. 1985,
%A\&A, 150, 53

%\bibitem De Young, D.S. \& Heckman, T.M. 1994, ApJ, 431, 598

\bibitem Edmunds, M.G., 1994, MNRAS, 270, L37

\bibitem Edvardsson, B., Anderson, J., Gustafson, B.,
Lambert, D.L., Nissen, P.E., \& Tomkin, J. 1993, A \& A, 275, 101

\bibitem Elbaz, D., Arnaud, M., \& Vangioni-Flam, E. 1995, A \& A, 303, 345

\bibitem Fields, B.D. \& Olive, K.A. 1996, Phys Lett B368, 103

\bibitem Fields, B.D., Kainulainen, K., Olive, K.A., \& Thomas, D. 1996
New Astronomy, in press

\bibitem Francois, P., Vangioni-Flam, E., \& Audouze, J. 1990, ApJ 361, 487

\bibitem Galli, D., Palla, F. Ferrini, F., \& Penco,U. 1995, ApJ,
433, 536

\bibitem Geiss, J. 1993, in {\it Origin
 and Evolution of the Elements} eds. N. Prantzos,
E. Vangioni-Flam, and M. Cass\'{e}
(Cambridge:Cambridge University Press), p. 89

%\bibitem Geiss, J. \& Reeves, H. 1972, A \& A, 18,126

\bibitem Gloeckler, G. \& Geiss, J. 1996, Nature, 381, 210

%\bibitem Grenon, M. 1989, Astrophys. \& Science, 156, 29

%\bibitem Grenon, M. 1990, in {\it Astrophysical Ages and Dating Methods},
%ed. E. Vangioni-Flam et al. (Ed. Frontieres, Paris), p. 153

%\bibitem Gry, C., Malinie, G., Audouze, J.,
%\& Vidal-Madjar, A. 1984, in
%Formation and Evolution of Galaxies and Large
%Scale Structure in the Universe,
%eds. J. Audouze \& J. Tran Tranh Van (Reidel, Dordrecht) p 279

\bibitem Hartwick, F.D.A. 1976, ApJ, 209, 418

\bibitem Hogan, C.J. 1995, ApJ, 441, L17

%\bibitem Iben, I. 1967, ApJ, 147, 624, 650

\bibitem Iben, I. \& Truran, J.W. 1978, ApJ, 220,980

\bibitem Iben, I. \& Tutukov, A. 1984, ApJ Suppl, 54, 335

\bibitem Larson, R.B. 1972, Nature, 236, 21

\bibitem Larson, R.B. 1974, MNRAS, 169, 229

\bibitem Larson, R.B. 1986, MNRAS, 218, 409

%\bibitem Lattimer, J., Schramm, D.N., \& Grossman, L. 1977, ApJ, 214, 819

\bibitem Linsky, J.L., Brown, A., Gayley, K., Diplas, A., Savage, B. D.,
Ayres, T. R., Landsman, W., Shore, S. N., Heap, S. R. 1993, ApJ, 402, 694

\bibitem Linsky, J.L.,  Diplas, A., Wood, B.E.,  Brown, A.,
Ayres, T. R.,  Savage, B. D., 1995, ApJ, 451, 335

\bibitem Loewenstein, M. \& Mushotzky, R. 1996, ApJ, 466, 695

%\bibitem Maeder, A. 1992, A \& A  264, 105

\bibitem Matteucci, F. \& Fran\c{c}ois, P. 1989, MNRAS, 239, 885

%\bibitem Molaro, P., Primas, F., \& Bonifacio, P. 1995, A \& A, 295, L47

%\bibitem Murphy, E.M., Lockman, F.J. \& Savage, B.D. 1995, ApJ, 447, 642

\bibitem Mushotzky, R. \etal 1996, ApJ, 466, 686

\bibitem Niemann, H.B. \etal 1996, Science, 272,846

%\bibitem Nissen, P.E. 1995, in IAU Symposium 164, {\it
%Stellar Populations}, to be published

\bibitem Olive, K.A. 1986, ApJ 309, 210

\bibitem Olive, K.A., Schramm, D.N., Scully, S. \& Truran, J.W. 1996,
in preparation

\bibitem Olive, K.A., Rood, R.T., Schramm, D.N., Truran, J.W.,
\& Vangioni-Flam, E. 1995, ApJ, 444, 680

%\bibitem Olive, K.A., \& Schramm, D.N. 1981, ApJ, 257, 276

\bibitem Olive, K.A., \& Scully, S.T. 1996, Int. J. Mod. Phys. A11,
409

\bibitem Olive, K.A., \& Steigman, G. 1995, ApJ S, 97, 49

\bibitem Olive, K.A., Thielemann, F.-K., \& Truran, J.W. 1987,
ApJ, 313, 813
 
\bibitem Ostriker, J.P., \& Thuan, T.X. 1975, ApJ, 202, 353

%\bibitem Ostriker, J.P., \& Tinsley, B. 1975, ApJ, 201, L51

\bibitem Pagel, B.E.J. 1988, in {\it Evolutionary Phenomena in Galaxies},
ed. J. Beckman \& B.E.J. Pagel (Cambridge University Press, Cambridge)

\bibitem Pagel, B E.J., Simonson, E.A., Terlevich, R.J.
\& Edmunds, M. 1992, MNRAS, 255, 325

\bibitem Rana, N.C. 1991,  ARA\&A, 29, 129

%\bibitem Reeves, H. 1978, in {\it Protostars and Planets},
%ed. T. Gehrels
%(Tucson:University of Arizona Press)

\bibitem Reeves, H. 1991, A\& A, 244, 294

\bibitem Rood, R.T., Bania, T.M., \& Wilson, T.L. 1992, Nature, 355, 618

\bibitem Rood, R.T., Bania, T.M.,  Wilson, T.L., \& Balser, D.S. 1995, 
in {\it
 the Light Element Abundances, Proceedings of the ESO/EIPC Workshop},
ed. P. Crane, (Berlin:Springer), p. 201

%\bibitem Rood, R.T., Steigman, G. \& Tinsley, B.M. 1976, ApJ, 207, L57

\bibitem Rugers, M. \& Hogan, C. 1996a, ApJ,  259, L1

\bibitem Rugers, M. \& Hogan, C. 1996b, astro-ph/9603084

\bibitem Rugers, M. \& Hogan, C. 1996c, in {\it Cosmic Abundances},
proceedings of the 6th Annual October Astrophysics Conference in Maryland,
PASP conference series, in press.

\bibitem Scalo, J. 1986, Fund. Cosm. Phys. 11, 1

\bibitem Schaller, G., Schaerer, D., Meynet, G., \& Maeder, A. 1992,
A\&AS, 96, 269

\bibitem Schatzman, E., 1987, A\&A, 172, 1

\bibitem Scully, S.T. \& Olive, K.A. 1995, ApJ, 446, 272

\bibitem Scully, S.T., Cass\'{e}, M., Olive, K.A., Schramm, D.N., 
Truran, J., and Vangioni-Flam, E. 1996, ApJ, 462, 960

%\bibitem Searle, L. \& Sargent, W.L. 1972, ApJ, 173, 25

%\bibitem Skillman, E., {\it et al.} 1995, ApJ Lett (in preparation)

%\bibitem Sommer-Larsen, J. 1991, MNRAS, 249,368

\bibitem Songaila, A., Cowie, L.L., Hogan, C. \& Rugers, M. 1994
Nature, 368, 599

\bibitem Spite, F. \&  Spite, M. 1982,  A.A., 115, 357

%\bibitem Steigman, G., Fields, B. D., Olive, K. A., Schramm, D. N.,
%\& Walker, T. P., 1993, ApJ 415, L35

%\bibitem Steigman, G. \& Tosi, M. 1992, ApJ, 401, 150

\bibitem Talbot, R.J. \& Arnett, W.D. 1973, ApJ 186, 69

\bibitem Tammann, G.A. 1994, in in {\it Supernova, Les Houches
Summer School Proceedings, Vol. 54}, ed. S. Bludman, R. Mochkovitch, \&
J. Zinn-Justin (Geneva: Elsevier Science Publishers), p. 1

\bibitem Thielemann, F.-K., Nomoto, K. \& Yokoi, K. 1986
A\&A, 158, 17

\bibitem Thorburn, J.A., 1994, ApJ, 421, 318

%\bibitem Timmes, F.X., \& Truran, J.W. 1995, preprint

\bibitem Timmes, F.X., Woosley, S.E., \& Weaver, T.A. 1995, ApJSupp,
98, 617

\bibitem Tinsley, B.M. 1980, Fund. Cosmic Phys., 5, 287

%\bibitem Tosi, M. 1988, A\&A, 197, 33

\bibitem Truran, J.W., \& Cameron, A.G.W. 1971, ApSpSci, 14, 179

%\bibitem Turck-Chi\`eze, S., Cahen, S., Cass\'e, M., \&
%Doom, C. 1988, ApJ, 335, 415

\bibitem Turner, M.S., Truran, J.W., Schramm, D.N., \& Copi, C.J. 1996,
astro-ph/9602050

\bibitem Tytler, D., Fan, X.-M., and Burles, S. 1996, astro-ph/9603069

\bibitem Vader, P. 1986, ApJ, 305, 669

\bibitem Vangioni-Flam, E., \& Audouze, J. 1988,
A\&A, 193, 81

\bibitem Vangioni-Flam, E. \& Cass\'{e}, M. 1995, ApJ, 441, 471

\bibitem Vangioni-Flam, E., Olive, K.A., \& Prantzos, N. 1994,
ApJ, 427, 618

\bibitem Vassiladis, E. \& Wood, P.R. 1993, ApJ, 413, 641

%\bibitem Vidal-Madjar, A. 1991, Adv. Space Res., 11, 97

\bibitem Walker, T. P., Steigman, G., Schramm, D. N., Olive, K. A.,
\& Kang, H. 1991 ApJ, 376, 51

\bibitem Wampler, E.J. \etal. 1996, astro-ph/9512084, AA, in press

\bibitem Wang, B. \& Silk, J. 1993, ApJ 406, 580

%\bibitem Wang, Q.D., Walterbos, R.A.M., Steakley, M.F., Norman, C.A.,
%\& Braun, R. 1995, ApJ, 439, 176

\bibitem Wasserburg, G.J., Boothroyd, A.I., \& Sackmann, I.-J. 1995, ApJ,
447, L37

\bibitem Wiedemann, V. 1977, A \& A, 61, L27

%\bibitem Weiler, R., Anders, E., Bauer, H., Lewis, R.
%\& Signer, P. 1991, Geochim \& Cosmochim Acta, 55, 1709

\bibitem Weiss, A., Wagenhuber, J., and Denissenkov, P. 1995, astro-ph/9512120

%\bibitem Woosley, S.E. \& Weaver, T.A. 1994, in {\it Supernova, Les Houches
%Summer School Proceedings, Vol. 54}, ed. S. Bludman, R. Mochkovitch, \&
%J. Zinn-Justin (Geneva: Elsevier Science Publishers), p. 100

\bibitem Woosley, S.E. \& Weaver, T.A. 1995, ApJ Supp, 101, 55

%\bibitem Wyse, R., \& Gilmore, G. 1995, AJ, 110, 2771

\bibitem Wyse, R., \& Silk, J. 1987, ApJ, 313, L11

\bibitem Yoshii, Y., Tsujimoto, T., \& Nomoto, K. 1996, ApJ, 462, 266

\endapjbib

\newpage   
       
\beginfig

\figitim {\bf Figure 1:} The evolution of D/H, and \he3/H (calculated
using the Iben \& Truran (1978) yields)
 for model Ia
with outflow (solid line) and without outflow (dotted line). The
evolution in model II is shown by the dashed line with outflow only.  Also shown
are the values of these ratios at the time of the formation of the sun and
today for D/H (open squares) and \he3/H (filled circles).

\figitim{\bf Figure 2:} The evolution of $^{16}$O/$^{16}$O$_\odot$ for model Ia
with outflow (solid line), without outflow (dotted line), and model II
(dashed line). 

\figitim{\bf Figure 3:} A comparison of the evolution of D/H and \he3/H,
as in Figure 1,
for models Ia, Ib, and Ic.

\figitim{\bf Figure 4:} A comparison of the evolution of 
$^{16}$O/$^{16}$O$_\odot$,
as in Figure 2, for models Ia, Ib, Ic.

\figitim{\bf Figure 5:} The time evolution of the outflow $o(t)$ for models 
Ia (solid), Ib (dotted), Ic (dot-dashed), and model II (dashed).

\figitim{\bf Figure 6:} The time evolution of the total (disk) mass of the
Galaxy relative to its initial mass for models
Ia (solid), Ib (dotted), Ic (dot-dashed), and model II (dashed).

\figitim {\bf Figure 7:} Distribution of G-dwarf stars as a function of
[Fe/H] - [Fe/H]$_1$ for the simple closed box model without the IRA
 (solid line), and with the IRA (dashed line) ([Fe/H]$_1$ is [Fe/H] at 
$t=14$ Gyr).
 Also plotted are the data taken from Pagel (1988).

\figitim {\bf Figure 8:} Differential distribution of G-dwarf stars as a
function of
[Fe/H] - [Fe/H]$_1$ for the simple closed box model without the IRA
 (solid line), and with the IRA (dashed line)
 plotted against the data taken from Pagel (1988).

\figitim {\bf Figure 9:} The age-metallicity relation showing
(Fe/H)/(Fe/H)$_\odot$ as
a function of time for the simple closed box models with (dashed)
 and without (solid) the IRA.

\figitim {\bf Figure 10:} As in figure 7, the 
distribution of G-dwarf stars as a function of
[Fe/H] for  model Ia (solid
line), model Ic (dot-dashed line), and model II (dashed line).
Model Ib looks very similar to Ic on this plot.

\figitim {\bf Figure 11:} As in figure 8, the
differential distribution of G-dwarf stars as a
function of
[Fe/H] for  model Ia (solid
line), 
model Ic (dot-dashed line) shown in a); and model II (dashed line) shown in b).

\figitim {\bf Figure 12:} The 
age-metallicity relation showing [Fe/H] as
a function of time for  model Ia (solid
line), model Ib (dotted line), 
model Ic (dot-dashed line), and model II (dashed line). 
The observational data taken from 
Edvardsson \etal (1993) is shown as a band.

\figitim{\bf Figure 13} Present day mass function of our model Ia as compared
with the data from Scalo (1986). 

\figitim{\bf Figure 14:} The evolution of D/H and \he3/H
 for model II
with outflow and the Iben \& Truran (1978) yields for \he3 (solid line)
 and in the same model with \he3 production in stars between 0.9 and 0.96
 M$_\odot$
as described in section 5 (dashed line).   

\endfig

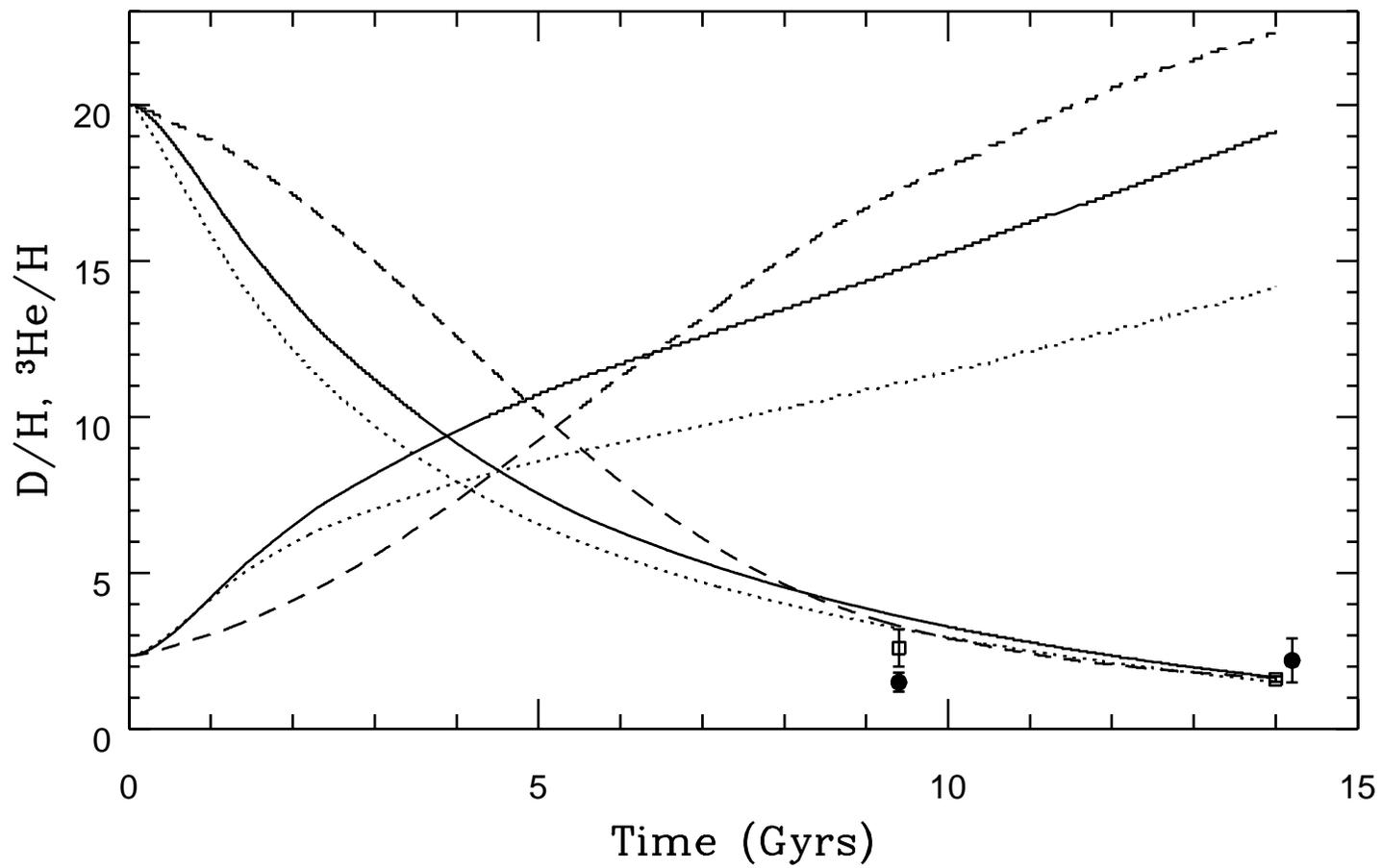
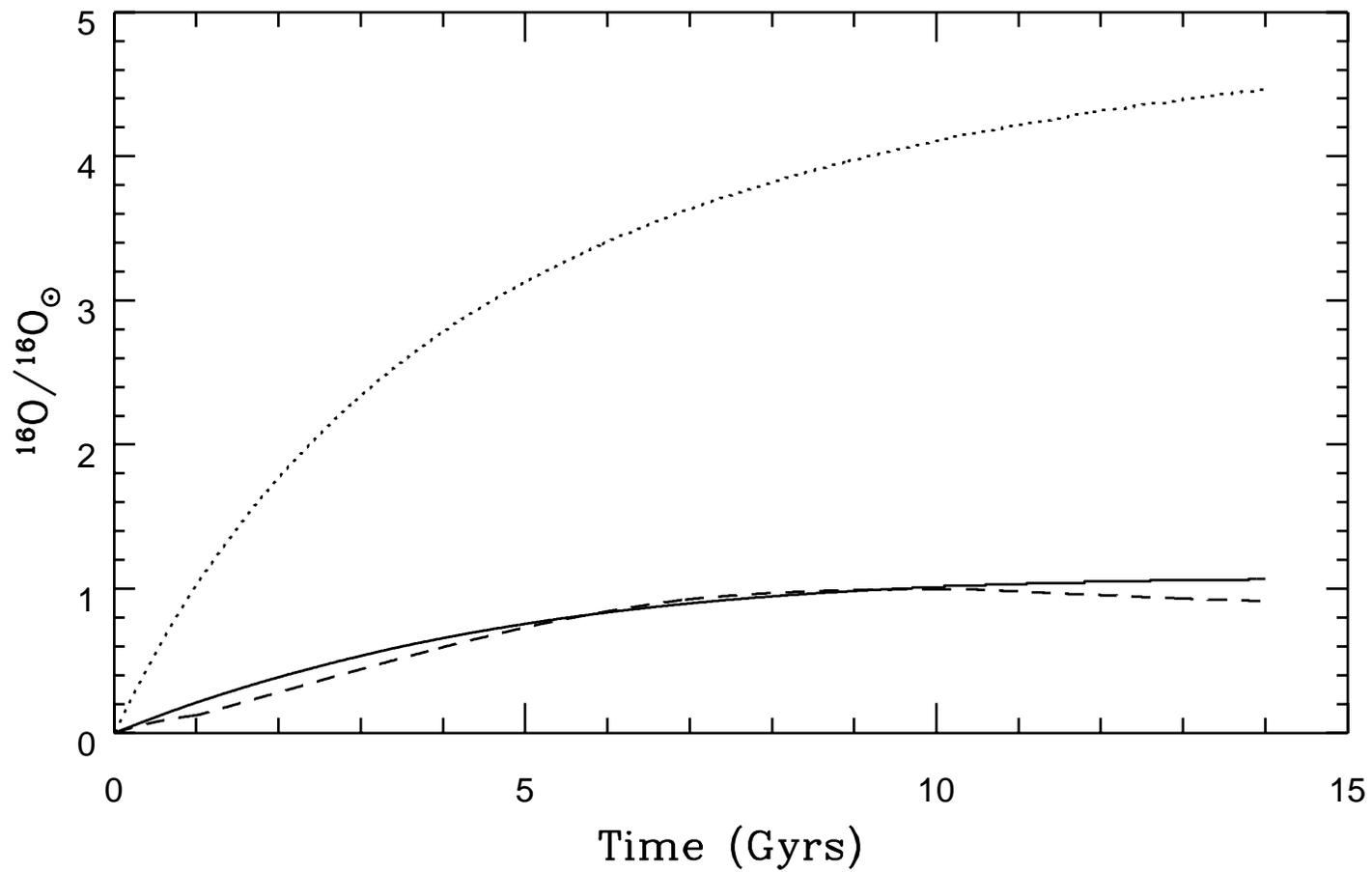
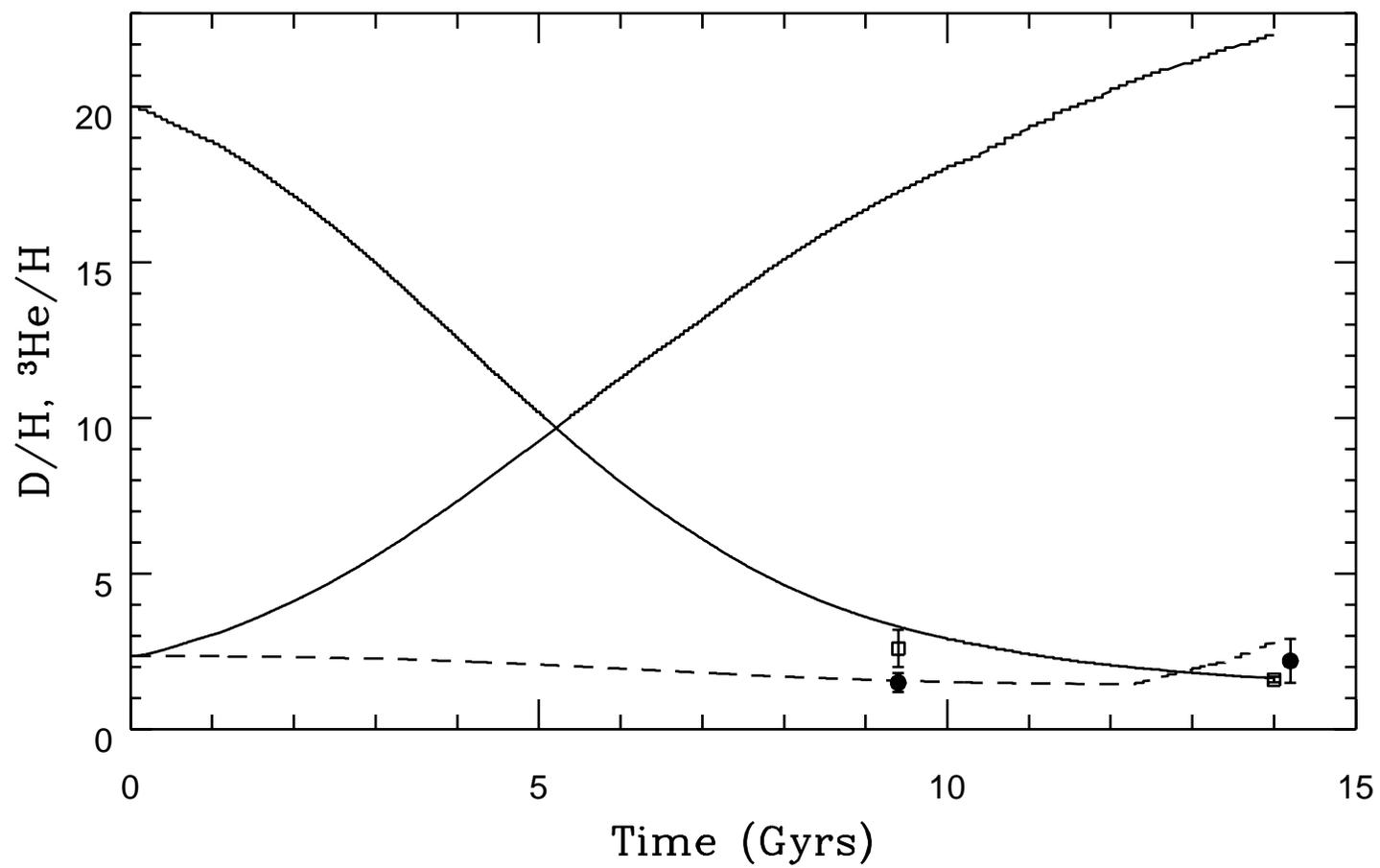
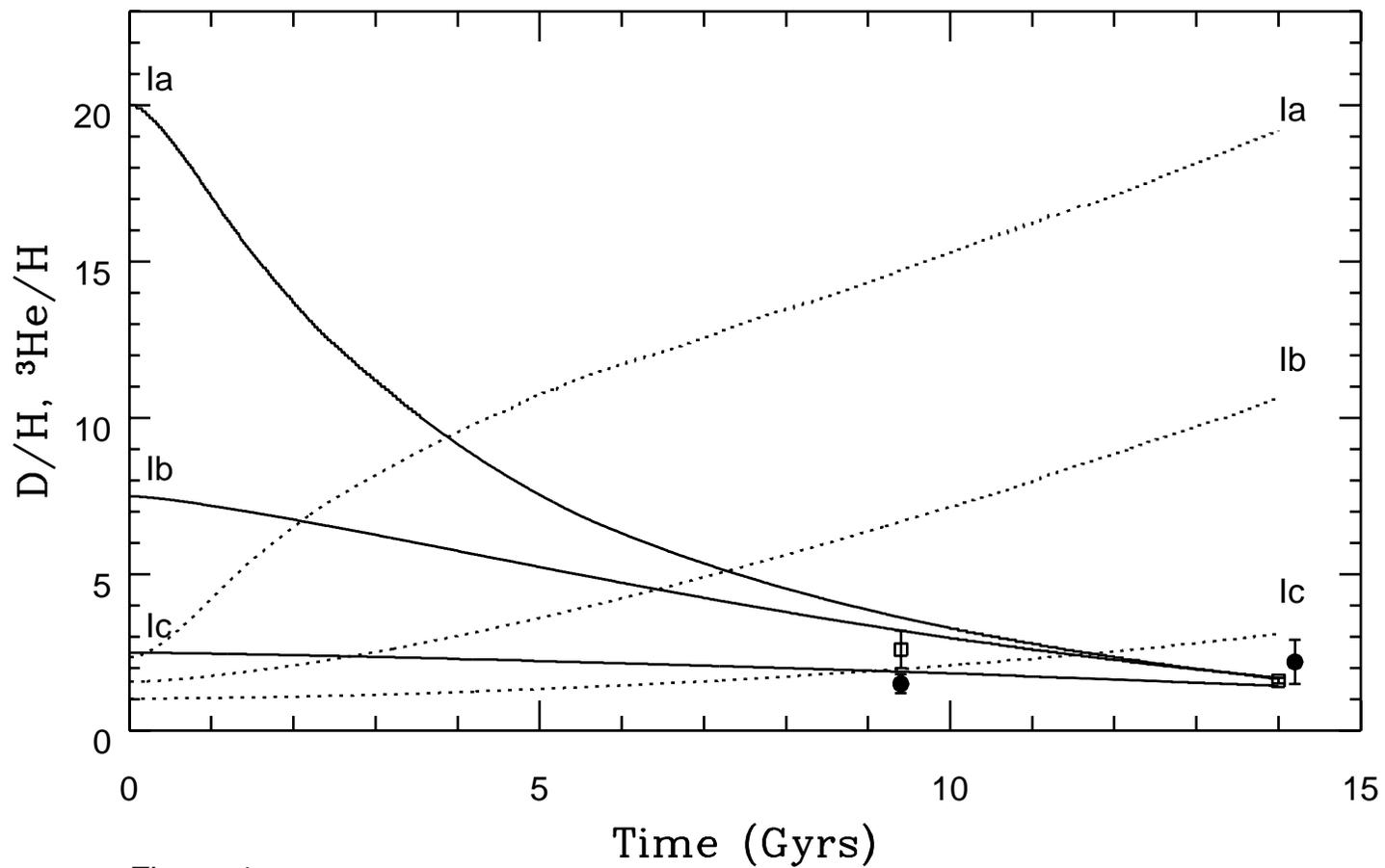
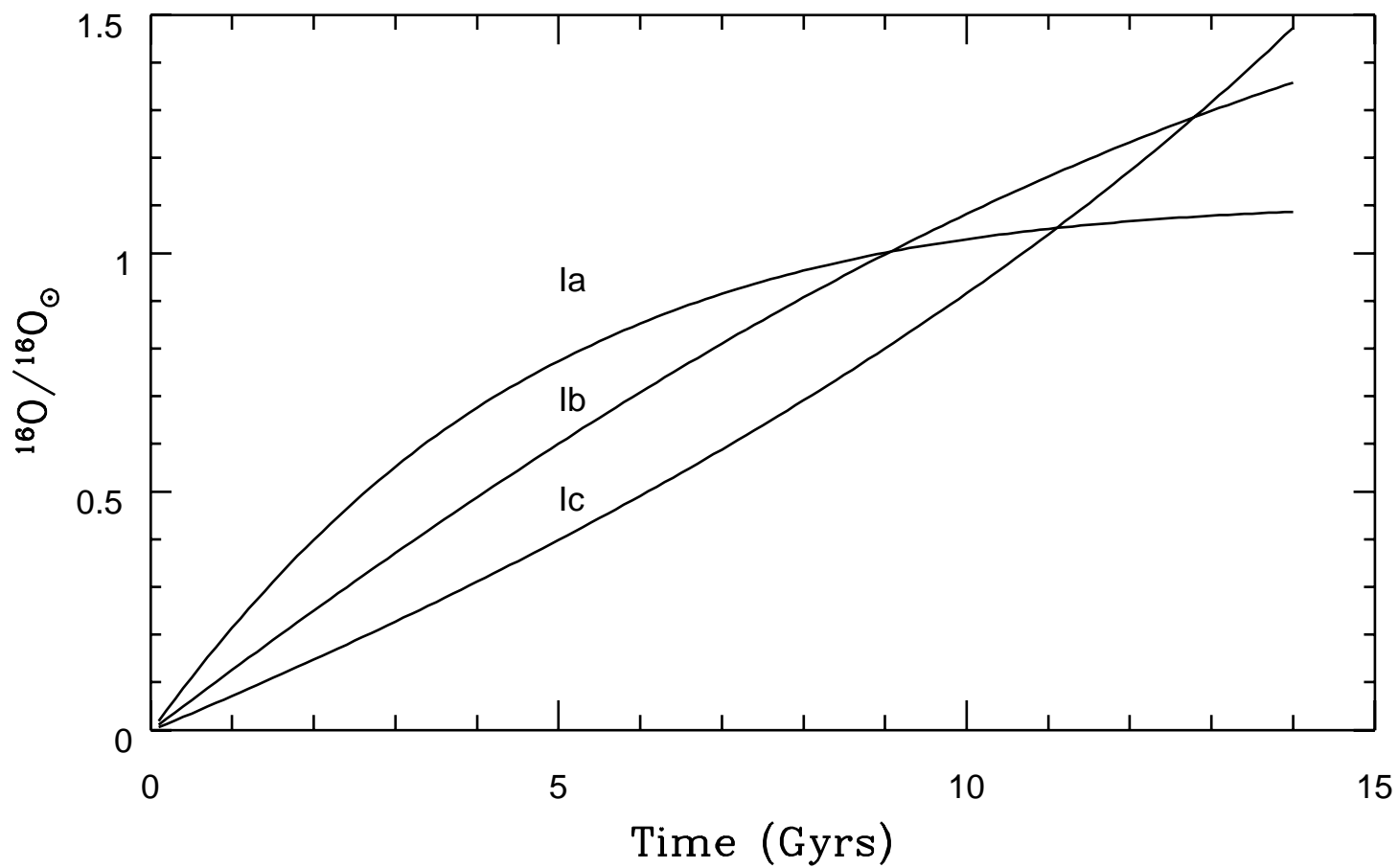
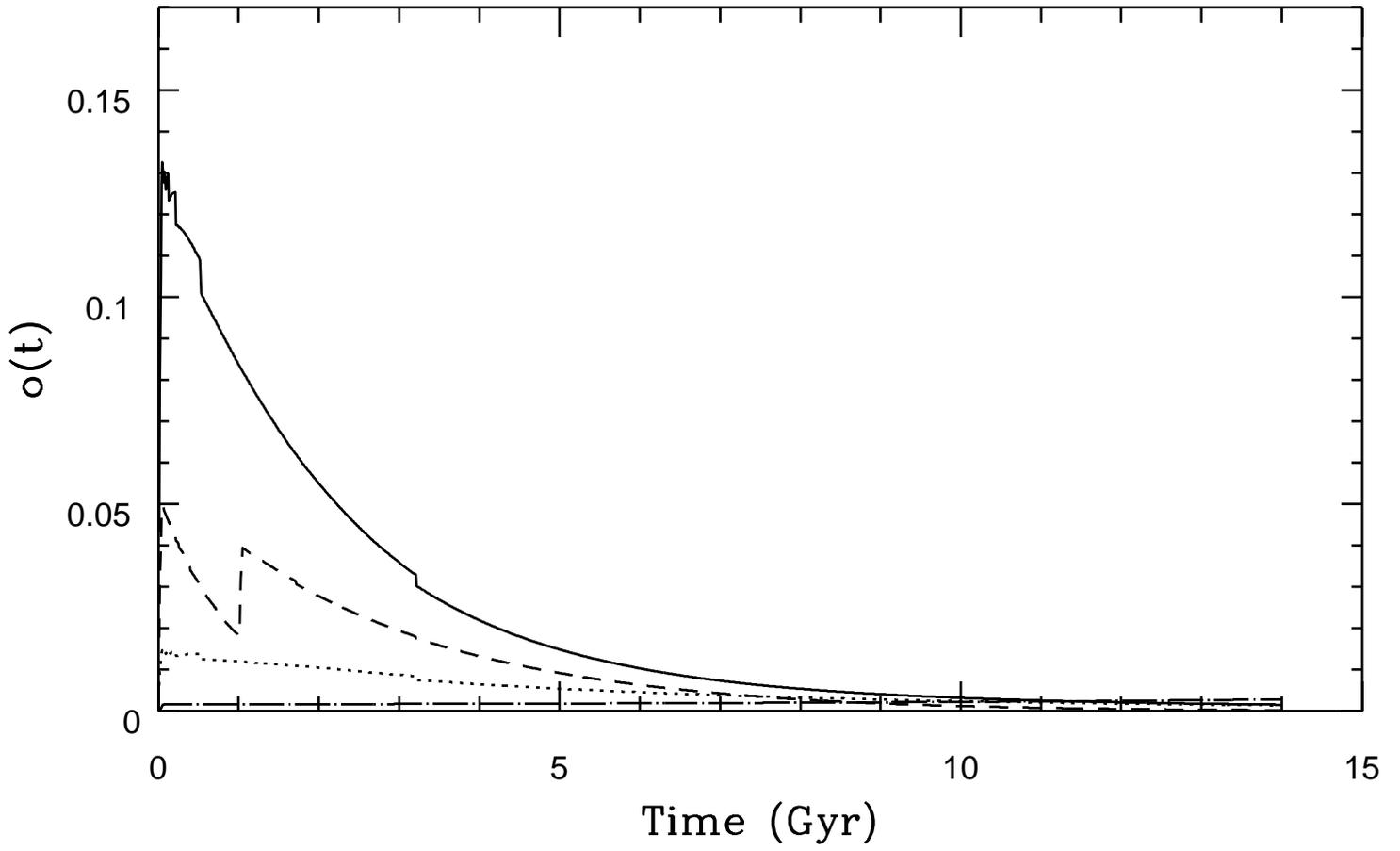
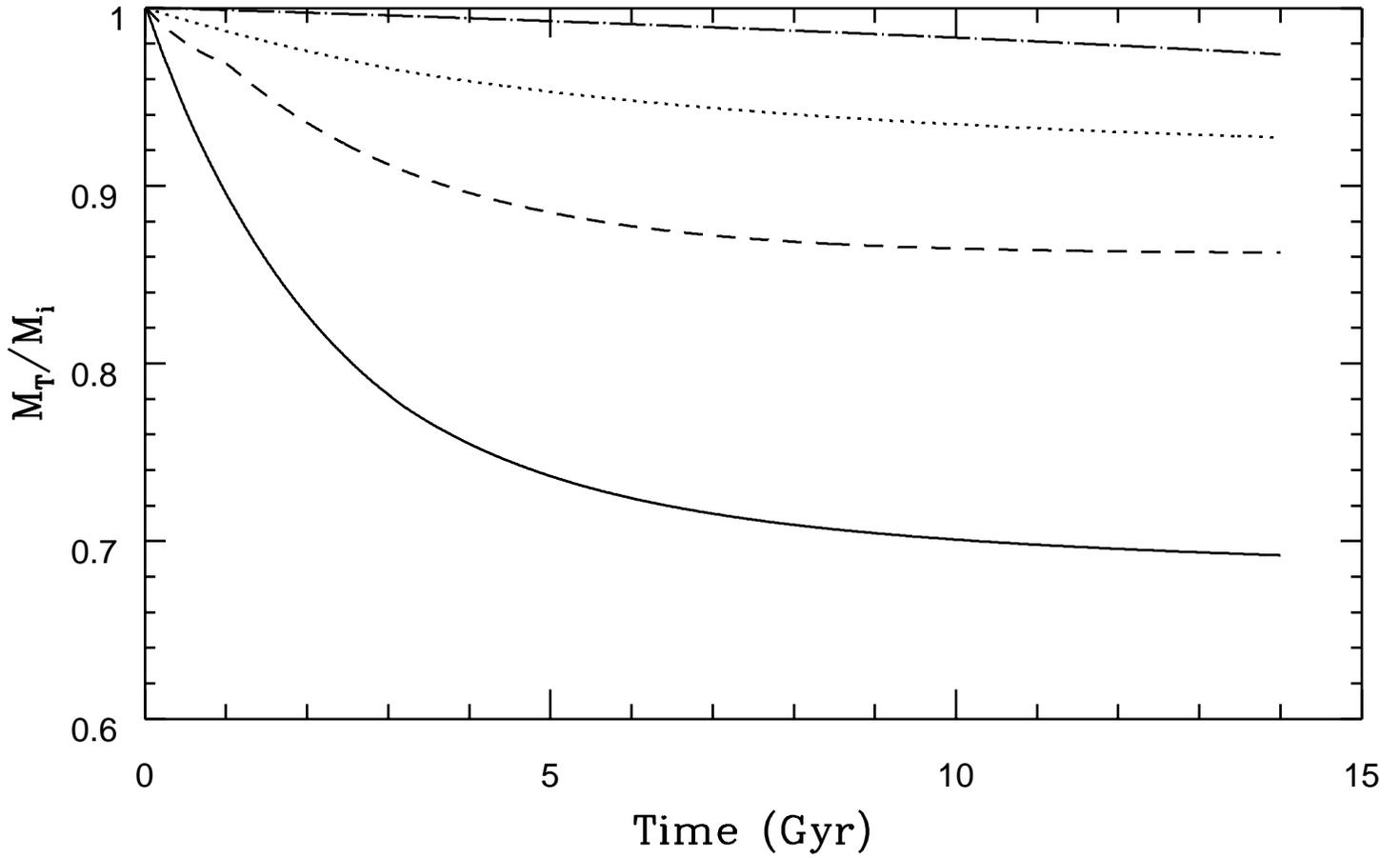
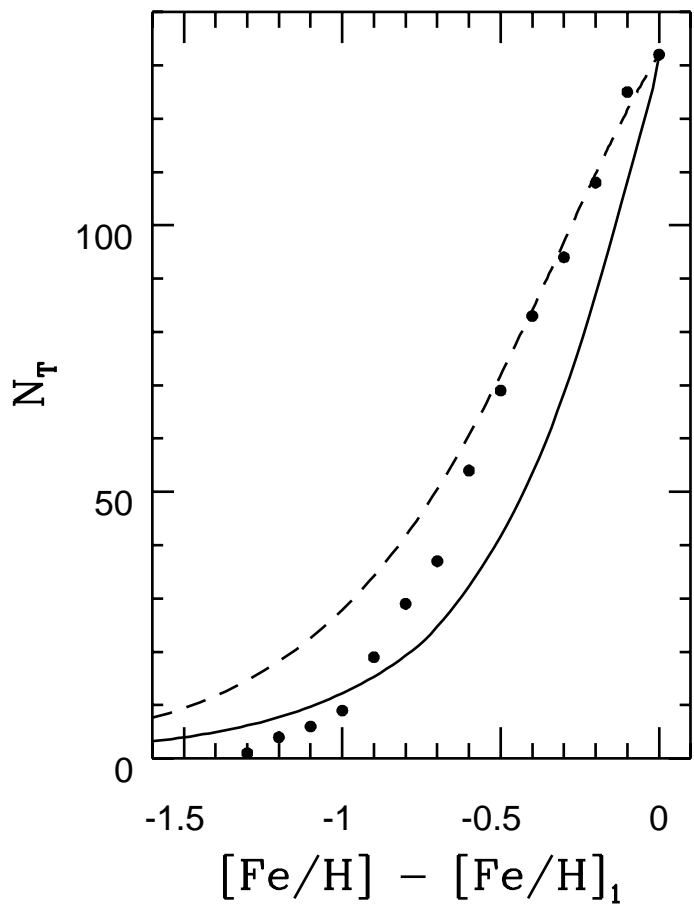
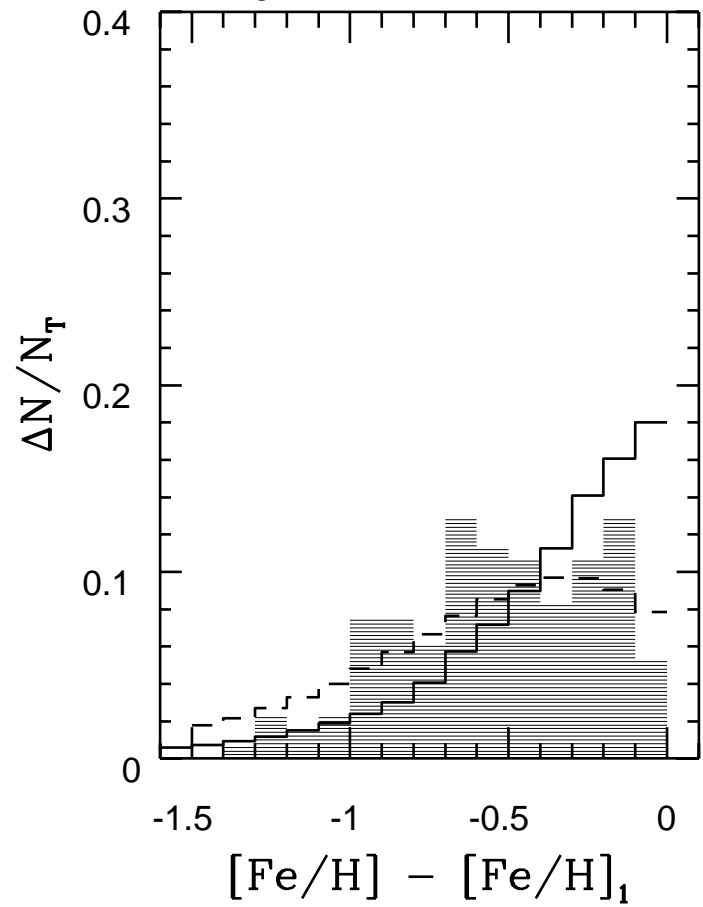
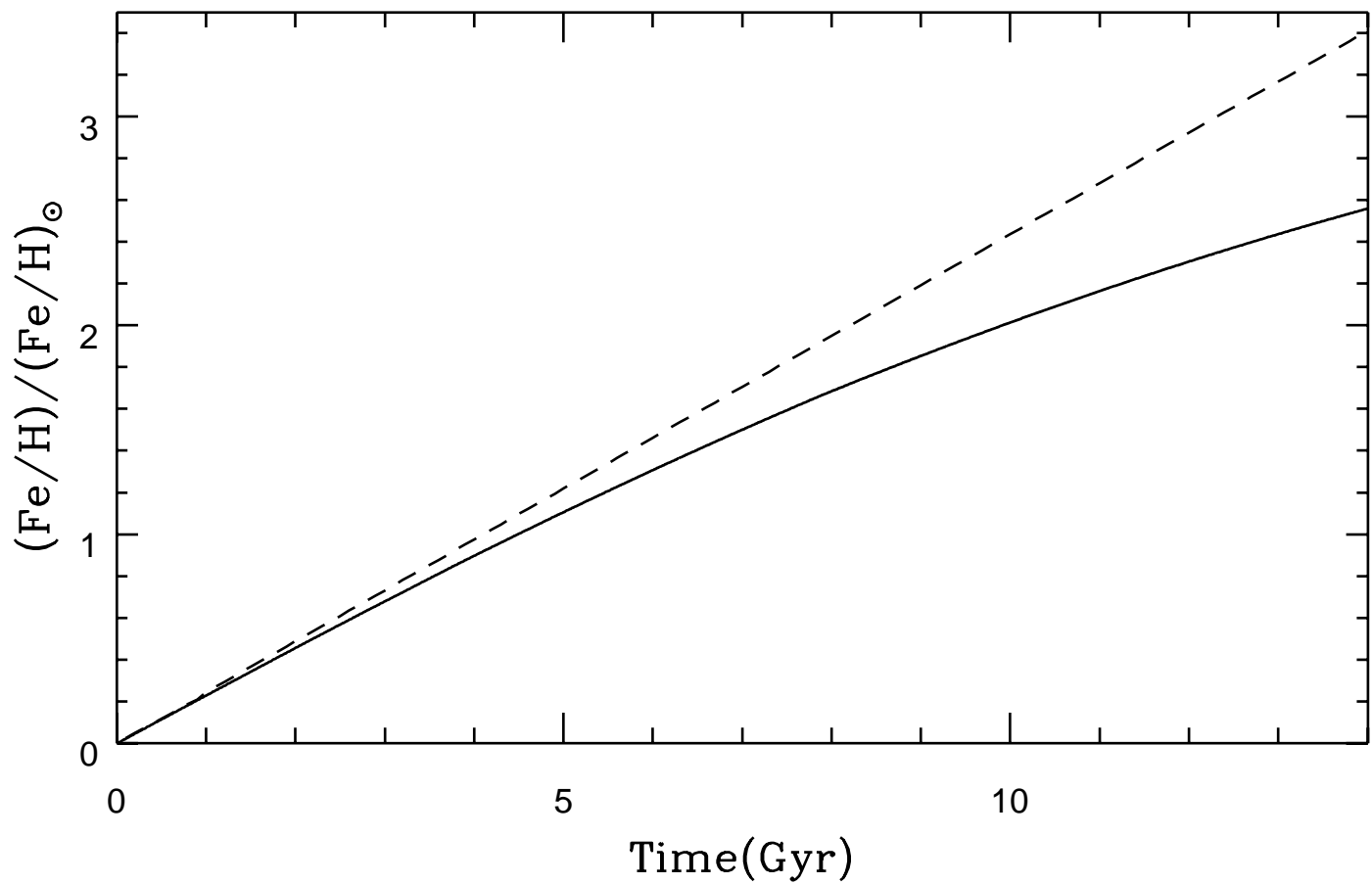
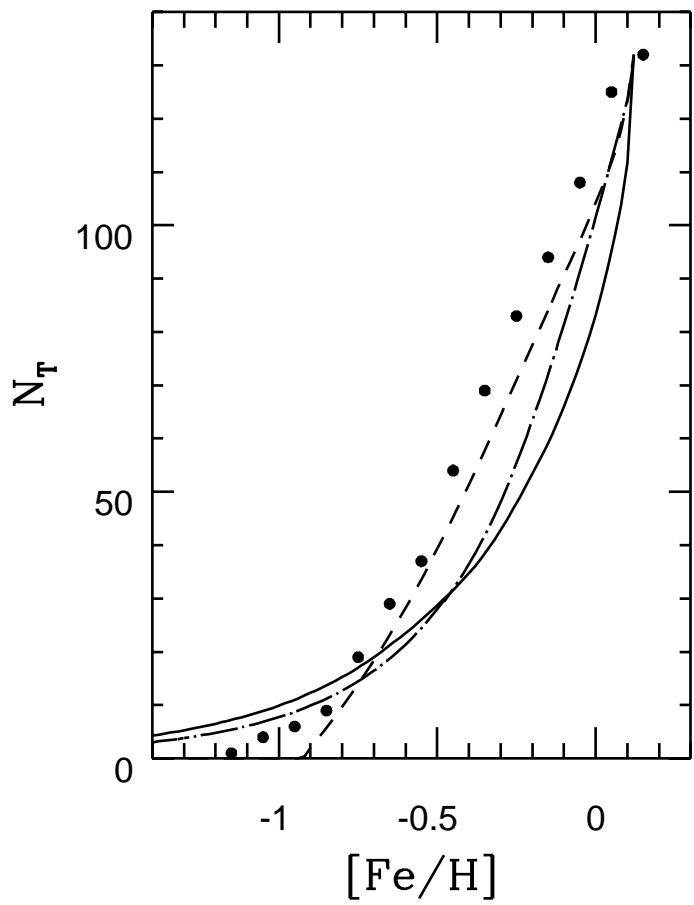
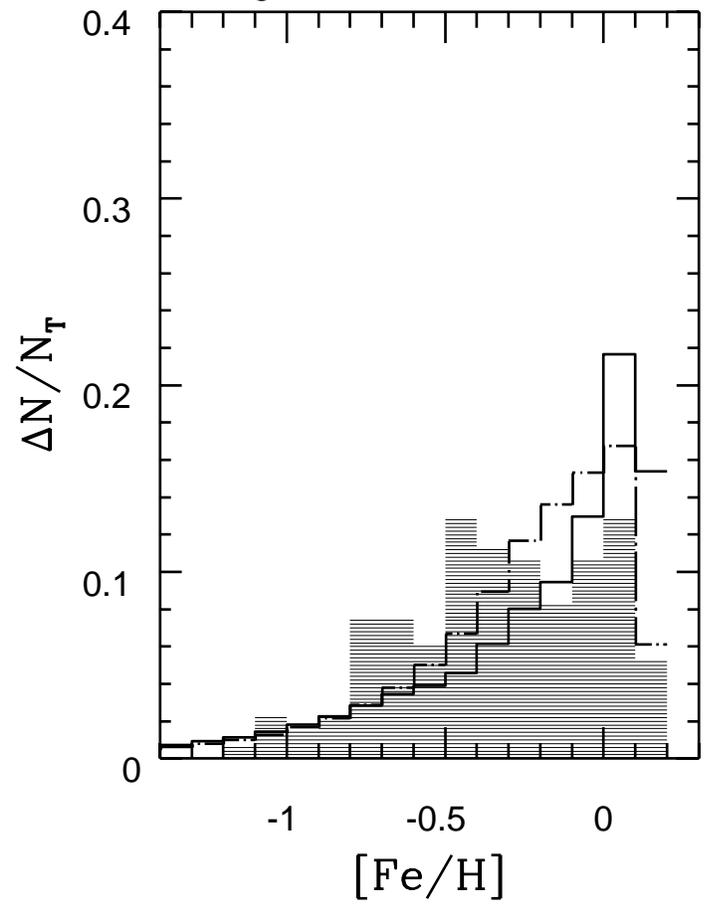
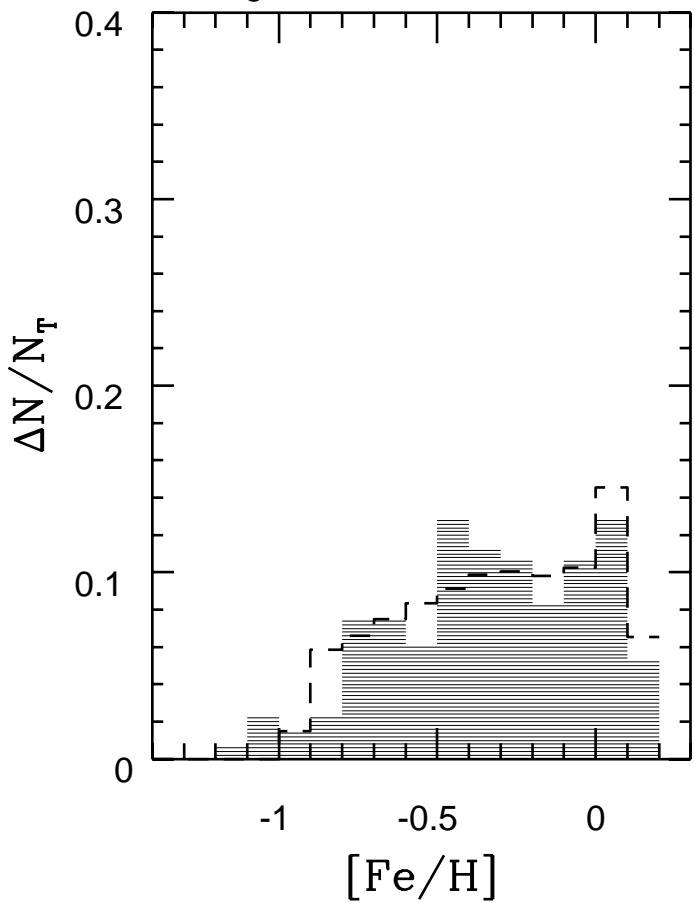
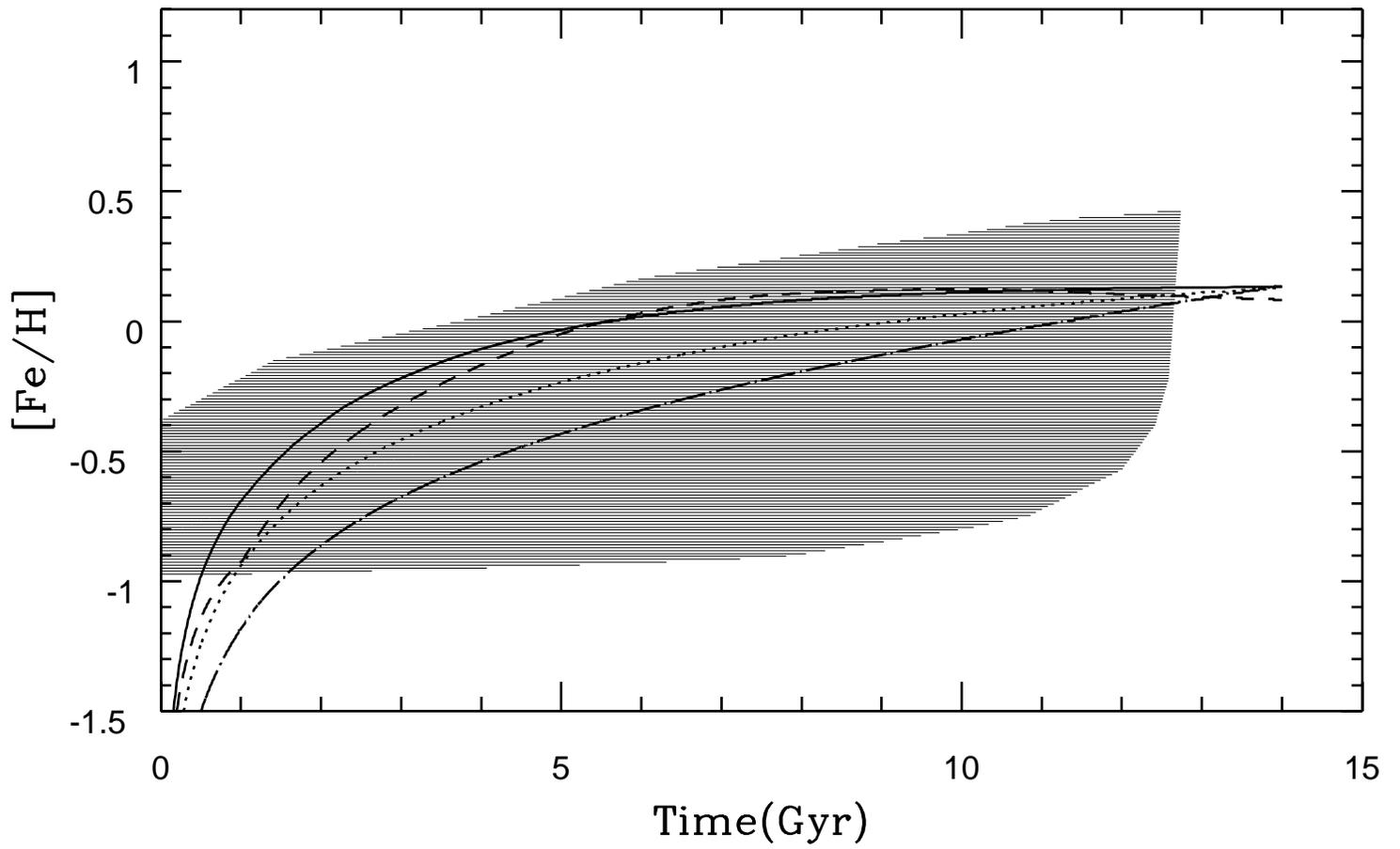
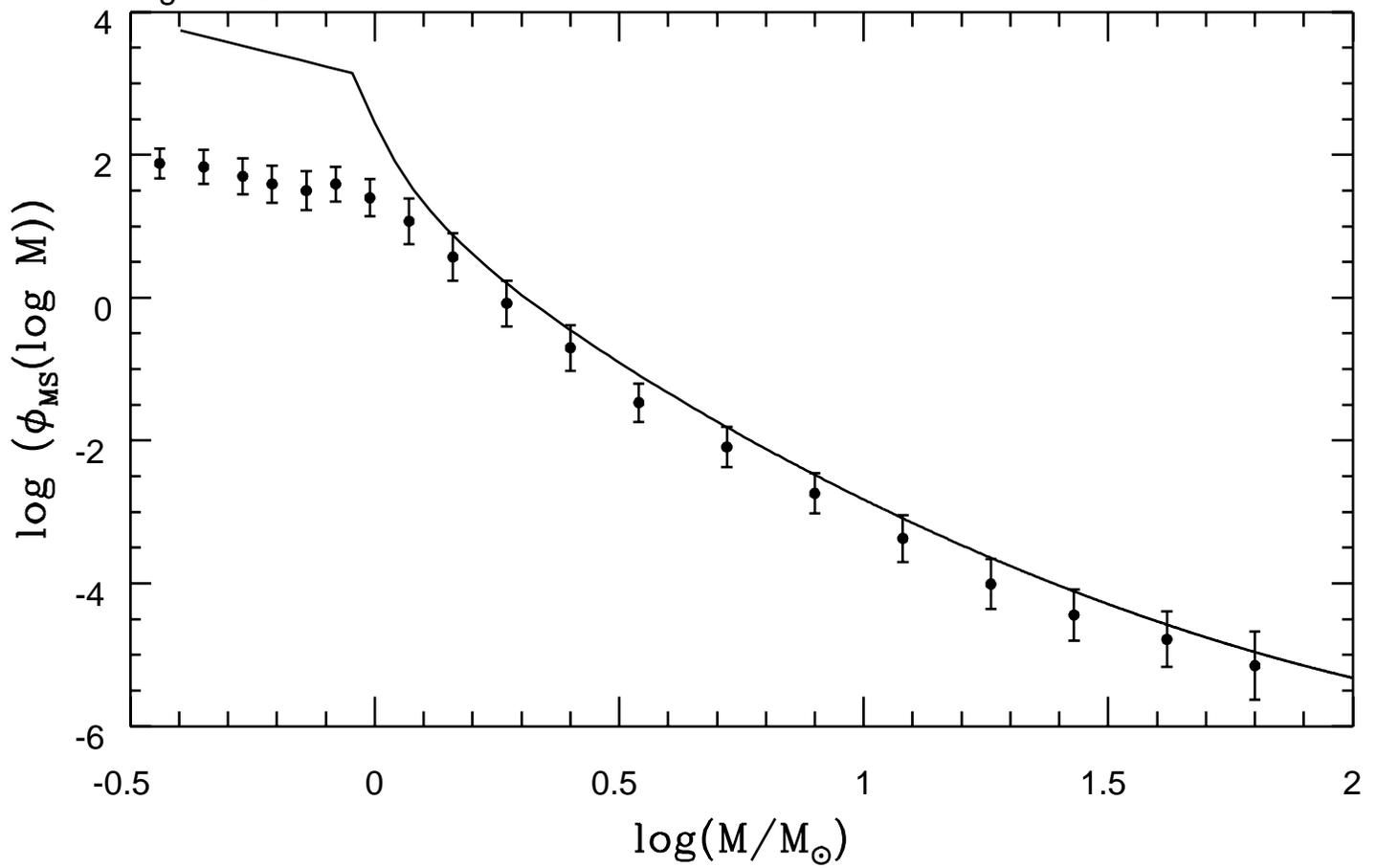
\end{document}